\documentclass[12pt]{article}

\usepackage{abstract}

\usepackage{algcompatible}
\usepackage{amsmath}
\usepackage{amssymb}
\usepackage{amsfonts}
\usepackage{amsthm}
\usepackage{array}
\usepackage[affil-it]{authblk}
\usepackage{booktabs}
\usepackage{bm}
\usepackage{color}
\usepackage{enumerate}
\usepackage{epsfig}
\usepackage{epstopdf}
\usepackage{eurosym}
\usepackage{float}
\usepackage{graphicx}
\usepackage{lineno}
\usepackage{listings}
\usepackage{lscape} 
\usepackage{mathtools}
\usepackage{mathrsfs}
\usepackage{multirow}
\usepackage[round]{natbib}
\usepackage{pdflscape}
\usepackage{pdfpages}
\usepackage{rotating}
\usepackage{setspace}
\usepackage{subfigure}
\usepackage{tablefootnote}
\usepackage{tabularx}
\usepackage{url}
\usepackage{verbatim}
\usepackage{etoolbox}
\AtBeginEnvironment{thebibliography}{\linespread{1}\selectfont}

\usepackage[ruled,vlined]{algorithm2e}
\usepackage{caption}
\usepackage[top=0.9in,bottom=0.9in,left=0.8in,right=0.8in]{geometry}
\usepackage[colorlinks=true,linkcolor=blue,citecolor=blue]{hyperref}
\usepackage[colorinlistoftodos]{todonotes}
\usepackage{listings}
\definecolor{mygreen}{rgb}{0,0.6,0}
\definecolor{mygray}{rgb}{0.5,0.5,0.5}
\definecolor{mymauve}{rgb}{0.58,0,0.82}
\lstdefinestyle{customc}{
	belowcaptionskip=1\baselineskip,
	breaklines=true,
	frame=L,
	xleftmargin=\parindent,
	language=C,
	showstringspaces=false,
	basicstyle=\footnotesize\ttfamily,
	keywordstyle=\bfseries\color{green!40!black},
	commentstyle=\itshape\color{gray},
	identifierstyle=\color{blue},
	stringstyle=\color{red},
}
\theoremstyle{definition}

\def\lsk{\left(}
\def\rsk{\right)}
\def\lbk{\left \{ }
\def\rbk{\right \} }
\def\lmk{\left [ }
\def\rmk{\right ] }

\DeclareMathOperator*{\supp}{\text{supp}}
\DeclareMathOperator*{\argmin}{argmin}

\newcommand{\mycircle}{\raise2pt\hbox{\textbullet}}

\newcommand{\vect}[1]{\boldsymbol{#1}}
\newcommand{\norm}[1]{\left\Vert#1\right\Vert}

\newcommand{\be}{\vect{e}}
\newcommand{\bbf}{\vect{f}}

\newcommand{\bu}{\vect{u}}
\newcommand{\bv}{\vect{v}}

\newcommand{\bx}{\vect{x}}
\newcommand{\by}{\vect{y}}
\newcommand{\bz}{\vect{z}}

\newcommand{\bF}{\vect{F}}

\newcommand{\bI}{\vect{I}}

\newcommand{\bX}{\vect{X}}

\newcommand{\bZ}{\vect{Z}}

\newcommand{\bbM}{\mathcal{M}}

\newcommand{\bbX}{\mathcal{X}}

\newcommand{\bbZ}{\mathcal{Z}}
\newcommand{\bbtZ}{\tilde{\mathcal{Z}}}

\newcommand{\bbU}{\mathbb{U}}

\newcommand{\bbeta}{\boldsymbol{\beta}}
\newcommand{\hbbeta}{\hat{\boldsymbol{\beta}}}

\newcommand{\bgamma}{\boldsymbol{\gamma}}

\newcommand{\bLambda}{\boldsymbol{\Lambda}}

\newcommand{\bepsilon}{\boldsymbol{\epsilon}}

\newcommand{\bzero}{\mathbf{0}}
\newcommand{\bone}{\mathbf{1}}

\title{\bf Variable Selection in Macroeconomic Forecasting with Many Predictors}

\author  {Zhenzhong Wang}
\author  {Zhengyuan Zhu}
\author  {Cindy Yu\thanks{Corresponding author: Cindy Yu, 2216 Snedecor, Iowa State University, Ames, IA 50010. Email: cindyyu@iastate.edu. Phone: 515-294-6885.}}
\affil{Department of Statistics, Iowa State University, Ames, IA}
\date{\vspace{-5ex}}

\begin{document}

\maketitle
\vspace{-1cm}
\begin{abstract}
{\bf Summary:} In the data-rich environment, using many economic predictors to forecast a few key variables has become a new trend in econometrics. The commonly used approach is factor augment (FA) approach. In this paper, we pursue another direction, variable selection (VS) approach, to handle high-dimensional predictors. VS is an active topic in statistics and computer science. However, it does not receive as much attention as FA in economics. This paper introduces several cutting-edge VS methods to economic forecasting, which includes: (1) classical greedy procedures; (2) $l_1$ regularization; (3) gradient descent with sparsification and (4) meta-heuristic algorithms. Comprehensive simulation studies are conducted to compare their variable selection accuracy and prediction performance under different scenarios. Among the reviewed methods, a meta-heuristic algorithm called sequential Monte Carlo algorithm performs the best. Surprisingly the classical forward selection is comparable to it and better than other more sophisticated algorithms. In addition, we apply these VS methods on economic forecasting and compare with the popular FA approach. It turns out for employment rate and CPI inflation, some VS methods can achieve considerable improvement over FA, and the selected predictors can be well explained by economic theories.
\end{abstract}

\begin{quotation}
\end{quotation}

\section{Introduction}

Recent development in information technology makes it possible to collect hundreds of economic variables in real time, with a reasonable cost. In such data-rich environment, using many economic predictors to forecast a few target variables has become a new trend in econometric research. In the last two decades, both theoretical and empirical works have been substantially built up on this direction, especially in the fields of macroeconomic forecasting \citep{SW2002a, BaiNg2008} and real-time now-casting \citep{GIANNONE2008,BANBURA2013}. This new trend has also made practical impact – economic forecasts using many predictors are currently being produced by fiscal and monetary authorities in both the U.S. \citep{FREDMD,FREDQD} and Europe \citep{ECB,REPEC2013}. A key aspect of many-predictor forecasts is to impose suitable parsimonious structure on data so that the curse of dimensionality is circumvented and useful information can be extracted. There are two directions to accomplish this, which are based on two different assumptions about the economic data structure.

The first direction is factor augment (FA) approach. It has been found to produce superior forecasts over traditional methods such as AR and VAR, thus attained favor from both econometricians and practitioners. This approach assumes that many predictors are relevant to the target variable and they have a factor structure. Dynamic factor model is applied first to compress the information of predictors into a handful of estimated factors. Then, the factors are augmented to a linear forecasting equation for the target variable. The rationale behind this approach is that the common variation among many observed economic variables can be represented by a handful of unobserved factors, and disturbances to these factors correspond to the major aggregate shocks to the economy such as demand or supply shocks \citep{SW2006}. This idea has a long tradition in macroeconomics
. One example is \cite{SW2002a}, which indicates the estimated factors can be interpreted as the diffusion indexes developed by NBER business cycle analysts to measure common movement in a set of macroeconomic variables.

Despite of the advantages of FA approach, there are a few drawbacks as well. First of all, it lacks explanation on the interrelationship among different economic variables, thus it cannot identify which predictors influence the target variable. Secondly, the estimated factors only capture the variation of major economic aggregates, but may lose information that is contained in a few predictors but beyond major economic aggregates. More importantly, the commonly used FA approach \citep{SW2002a,SW2002b} does not take into account the target variable when estimating the factors, which means the factors used in the forecasting equation are the same no matter which target variable is being forecasted. 

The second direction of many-predictor forecasts is to directly select the best predictors and their lagged values to carry out forecast, and we call it variable selection (VS) approach. This direction implies another rationale of the economic data -- given the selected predictors in the forecasting equation, all others have insignificant prediction power to the target variable anymore. To be noticed, it does not means the unpicked predictors are irrelevant or independent to the target variable. The forecasting equation derived from VS approach indicates that, conditional on the selected predictors, the remaining predictors have little prediction power on the target variable. VS is not a new topic, but it has not drawn as much attention as FA in economic forecasting. In contrast, VS has substantial development in other fields such as statistics, computer science, and bioinformatics, and impressive new methods and applications keep coming forward. 

The first goal of this paper is to review several cutting-edge VS methods, and compare their performance with FA approach in the context of economic forecasting. One advantage of VS is its capability of interpreting the individual impact of each predictor on the target variable, including both direction and magnitude. This is helpful for understanding the interrelationship among different economic variables. More importantly, VS can select predictors that may contain useful information beyond the aggregate economic activity explained by the factors in FA approach. Only including the important predictors will avoid overfitting issue, thus enhance prediction power. In our empirical studies in Section \ref{chap4:application}, we apply both FA and VS approaches to forecast three important macroeconomic variables -- Employment (EMP), Industrial Production (IP) and Consumer Price Index (CPI), and find that some VS methods achieve considerable improvement over FA approach for EMP and CPI. Also the relationship between the target variable and the selected predictor can be well explained by economic theories.  

The second goal of this paper is to evaluate several groups of VS methods, including both classical procedures and cutting-edge algorithms, in terms of both variable selection accuracy and out-of-sample forecasting. Due to the huge body of VS literature, it is impossible to do an exhaustive review for VS methodologies. For this paper, we only focus on the high dimensional regime (dimension of predictors is larger than the number of observations), which is the case of economic forecasting. We pick the following four groups of methods: (1) classical procedure (forward selection); (2) $l_1$ regularization (adaptive LASSO); (3) gradient descent algorithms with sparsification (iterative hard thresholding and thresholding pursuit); and (4) a meta-heuristic algorithm called sequential Monte Carlo (SMC) proposed by \cite{Duan2019}. We do not consider any machine learning algorithm such as random forest and neural network, due to their lack of interpretability. All these VS methods are applied in the framework of linear regression. Their performance in both variable selection and out-of-sample forecasting are examined through several simulation studies. The results show that, SMC, the most time-consuming algorithm, works the best across all simulation settings. Surprisingly, the performance of the classical forward selection matches up to the SMC and better than other advanced modern methods ($l_1$ regularization and gradient descent algorithms with sparsification). 

The remainder of the paper is structured as follows. Section \ref{chap4:ecoforecast} introduces the setting of economic forecasting with many predictors, and the implementation of FA and VS approaches. Section \ref{chap4:vsreview} briefly reviews the four groups of VS methods, including their methodologies, advantages and disadvantages. Several simulation studies are carried out in Section \ref{chap4:simulation} to evaluate the performance of the four groups of VS methods. In Section \ref{chap4:application} we apply these VS methods on economic forecasting using the FRED-MD database (\cite{FREDMD}), and compares their forecasting performance with that of FA approach. Conclusions and discussions are presented in Section \ref{chap4:conclusion}.

\paragraph{Notations} Throughout this paper, bold letters denote vectors, unbold letters denote scalar quantities and calligraphy letters denote matrices. $\bzero$ and $\bone$ stand for a vector of zeros and ones respectively. For a $p$-dimensional vector $\bv=(v_1,\dots,v_p)'$, we use $\norm{\bv}_0=\sum_{i=1}^p I(v_i\neq 0)$ with $I(.)$ being the indicator function, $\norm{\bv}_1=\sum_{i=1}^p|v_i|$ and $\norm{\bv}_2=\sqrt{\sum_{i=1}^p|v_i|^2}$ to denote the $l_0$ norm, $l_1$ norm and $l_2$ norm of $\bv$ respectively. $\supp(\bv)$ denotes the support of $\bv$, i.e. $\supp(\bv)=\{$indices of nonzero elements in $\bv\}$. For a set $U$, $|U|$ denotes its cardinality, i.e. the number of elements in $U$. $\bv_U=[v_i]_{i\in U}$ stands for a sub-vector of $\bv$ whose indices of elements belong to $U$. For every iterative algorithm, we use superscript $(r)$ to stand for the $r$-th iteration.

\section{Economic Forecasting with Many Predictors}
\label{chap4:ecoforecast}
In this section we first describe the setting of economic forecasting with many predictors, including notations and assumptions, then we outline FA and VS approaches under this framework.

\subsection{Setting}
\label{subchap4:setup}
We adopt the notations and assumptions per usual in economic forecasting literature \citep{SW2002a,SW2006,Ng2013}. Let $y^h_{t+h}$ be the $h$-step ahead value of the variable to be forecasted.  For example, in Section \ref{chap4:application} we consider forecasts of 1, 3, 6 and 12-month growth of the Employment (EMP). Let $EMP_t$ denote the value of EMP on month $t$. Then the $h$-month growth of EMP, at an annual rate, is
\begin{equation}
y^h_{t+h} = (1200/h)\log(EMP_{t+h}/EMP_{t}).
\label{eq:yemp}
\end{equation}
Let $\bZ_t$ be the $n$-dimensional vector of predictor variables, which also includes the current value of the target variable $y_t$. In economic forecasting, both $y^h_{t+h}$ and $\bZ_t$ are required to be stationary. This is accomplished by suitable preliminary transformations which are determined by a combination of statistical tests and expert judgment. For instance, unit root tests indicate that the logarithm of industry production (IP) series (denoted as $\{IP_t\}$) has a unit root. Therefore, its appropriate transformation is taking the log first difference, i.e. the corresponding predictor variable is $z_t=\log(IP_{t}) - \log(IP_{t-1})$. After transformation, each predictor $z_t$ is standardized to have mean zero and sample variance one. This standardization is required for FA approach and some VS methods. 

\subsubsection{FA Approach}
\label{subchap4:FA}
FA approach assumes the predictor variables admit the following factor model representation: 
\begin{equation}
\bZ_t = \bLambda \bF_t + \be_t,
\label{eq:factor}
\end{equation}
where $\bF_t$ is the $s \times 1$ latent factors, $\bLambda$ is the $n\times s$ matrix of factor loadings, and $\be_t$ is the vector of idiosyncratic components satisfying $E(\be_t|\bF_t) =\bzero$ and finite second moments. Here the latent factors $\bF_t$ are estimated by the principle component analysis. \cite{SW2002a} has proved that the principal component estimator is point-wise  (for any date $t$) consistent and has limiting mean squared error (MSE) over all $t$ that converges to 0, under a suitable set of identifiability conditions. If some series contain missing values, the expectation-maximization (EM) algorithm described in \cite{SW2002a} is utilized to estimate factors $\bF_t$. After the factors have been estimated, the $h$-step ahead forecast is the linear projection of $y^h_{t+h}$ onto the $t$-dated factors, $y_t$ and their lagged values:
\begin{equation}
y_{t+h}^h = \alpha + \sum_{l=0}^{q-1} \alpha_l y_{t-l} + \sum_{l=0}^{m-1} \bgamma_l'\bbf_{t-l} +  \varepsilon_{t+h}.
\label{eq:faforecast}
\end{equation}
Here $q$ is the auto-regressive order, $m$ is the order of lagged factors and $\bbf_t$ is a vector of first $d$ factors in $\bF_t$.
In practice, $q$, $m$ and $d$ can be selected by some criteria, such as Schwarz’s BIC \citep{schwarz1978} and forward cross validation (FCV).  

Note that factor model (\ref{eq:factor}) only includes the current value of predictor variables ($\bZ_t$) without considering their lagged values. The historical information of $\bZ_t$ are incorporated to forecasting through the lagged value of factors ($\bbf_{t-l}$, $l=1,\dots,m-1$). 
In our empirical study, we have 128 monthly economic time series, thus the vector $\bZ_t$ for FA approach is 128-dimensional. However, the predictors used in VS approach are $\bX_t=\lsk \bZ'_t,\bZ'_{t-1},\dots,\bZ'_{t-5}\rsk'$, assuming that the lag order is five. Then the dimension of $\bX_t$ is larger than the number of observations. In order to distinguish the predictors in these two different approaches, we use $\bZ_t$ to denote the predictor variables in FA approach, and use $\bX_t$ to denote the predictors in VS approach, respectively.

\subsubsection{VS Approach}
\label{subchap4:VS}
We apply various VS methods on economic forecasting through the following linear regression:
\begin{equation}
y_{t+h}^h = \beta_0 + \bX_t'\bbeta + \epsilon_{t+h}, \quad t=1,\dots,T,
\label{eq:vsforecast}
\end{equation} 
where $\bX_t=(X_{1t},\dots,X_{pt})'$ is a $p-$dimensional vector of the predictors. In our real data application, $p=6n=768$. and $\bbeta=(\beta_1,\dots,\beta_p)'$ is the vector of regression coefficients. The matrix form of (\ref{eq:vsforecast}) is as follows:
\begin{equation}
\by = \beta_0\bone + \bbX\bbeta + \bepsilon,
\label{eq:vsmat}
\end{equation} 
\vspace {-5mm}
\begin{equation}
\by=(y^h_{1+h},\dots,y^h_{T+h})', \quad \bbX=[\bX_1,\cdots,\bX_T]', \quad \bepsilon=(\epsilon_{1+h},\dots,\epsilon_{T+h})'.
\end{equation} 
Here we use $\bx_i$ ($i=1,\dots,p$) to denote each column of the model matrix $\bbX$, i.e. $\bbX=\lmk\bx_1,\dots,\bx_p \rmk$. As mentioned in Subsection \ref{subchap4:setup}, all $\bx_i$'s are standardized to have mean zero and sample variance one. 

The center part of the VS approach is the best subset problem with subset size $k$, which is given by the following optimization:
\begin{equation}
\min_{\bbeta}\;\norm{\by - \beta_0\bone - \bbX\bbeta}^2_2 \quad\text{subject to} \quad \norm{\bbeta}_0\leq k.
\label{eq:vsopt}
\end{equation}
Here the $l_0$ norm of $\bbeta$ (i.e. $\norm{\bbeta}_0$) counts the number of nonzeros in $\bbeta$, which is bounded by $k$. Let $\hbbeta_k$ be the optimal solution of (\ref{eq:vsopt}), then the support of $\hbbeta_k$, denoted as $\hat{U}_k:=\lbk i: \hat{\beta}_i\neq 0 \rbk$, is the best subset of predictors with size $k$. In practice, the subset size $k$ can be determined by AIC \citep{Akaike1974}, BIC, FCV or other criteria. 

The discrete nature of cardinality constraint ($\norm{\bbeta}_0\leq k$) poses a great difficulty in finding the global optimum. It requires comparison of all $\binom{p}{k}$ subsets of predictors, which is infeasible for large $p$. To the best of our knowledge, computing the optimal solution to problem (\ref{eq:vsopt}) is in general deemed as intractable. However, the last few decades have seen a flurry of activity in developing algorithms trying to solve (\ref{eq:vsopt}) at reasonable time cost, with associated optimality under certain conditions. In Section {\ref{chap4:vsreview}}, we will review four groups of VS methods that try to obtain the good sub-optimal solution more efficiently. 

\section{Overview of Variable Selection Methods}
\label{chap4:vsreview}

As there is a vast literature on this topic, we present a selective overview. We select the following four types of VS methods: (1) classical greedy procedures, (2) $l_1$ regularization methods, (3) gradient descent algorithms with sparsification, and (4) meta-heuristic algorithms. The first two groups have already been investigated in many econometric literature, therefore their introduction are relatively concise. The last two groups are proposed in computer science and mathematical optimization but have not been seen wide adoption in economic forecasting. Thus these two groups will be introduced more elaborately. For each group, we mainly focus on the methods applied in our empirical study, presenting their ideas, advantages and disadvantages. The algorithm details can be found in Appendix \ref{sec:appB}. 

\subsection{ Classical Greedy Procedures}
Classical VS procedures such as forward selection (FS), backward elimination (BE), and stepwise regression (SR) are available in many statistical software packages. These algorithms are greedy algorithms, which follow the heuristic of making the locally optimal choice at each iteration with the intent of finding a global optimum. For example, when adding a new predictor to the model, FS selects the one which maximize the decrement of sum of square errors (SSE) given the predictors already included in the model, until the model has $k$ predictors in total. As a result, such greedy search only examines a small portion of possible subsets of predictors and may be trapped in a local solution thus cannot guarantee to obtain the global optimum. 

Due to the nature of greedy algorithms, the classical procedures seem to be inferior to modern VS algorithms such as LASSO. However, the latter also rely on certain assumptions to achieve global optimality, and these assumptions are hard to be verified in practice. In addition, FS and BE do not involve any tuning parameters, while more advanced algorithms are usually sensitive to the choice of tuning parameters and initial values. Thus, it is meaningful to compare classical greedy procedures with the modern algorithms in terms of their empirical performance. Moreover, another advantage of FS and BS is their fast update of model estimation when adding or deleting a predictor, which can avoid calculating the inverse of $\bbX'\bbX$. This fast updating algorithm and the detailed procedures of FS and BE are presented in Appendix \ref{subsec:appclassical}. 

\subsection{$l_1$ Regularization Methods}
Since it is hard to directly solve the $l_0$ constraint optimization (\ref{eq:vsopt}), a group of methods pursue another direction of VS -- $l_1$ regularization, which is a convex relaxation of the $l_0$ constraint. One representative is LASSO \citep{Tibshirani1996}:
\begin{equation}
\min_{\bbeta}\;\lbk\norm{\by - \beta_0\bone - \bbX\bbeta}^2_2+\lambda\norm{\bbeta}_1\rbk.
\label{eq:lasso}
\end{equation}
As a surrogate for problem (\ref{eq:vsopt}), $l_1$ regularized methods produce a sparse estimation by shrinking many coefficients toward zero. There has been a large amount of work on this topic in terms of algorithms, theoretical properties and real world applications. Readers can refer to the books \cite{buhlmann2011statistics}, \cite{hastie2009elements}, \cite{wainwright_2019} and references therein. $l_1$ regularization methods enjoy several attractive properties. The first advantage of $l_1$ regularization, which is also an important reason to its popularity, is great computational efficiency. The problem (\ref{eq:lasso}) is a convex quadratic optimization and there are several efficient algorithms to solve it. For example, 
pathwise coordinate optimization \citep{friedman2007} can compute the solution path at the same cost as a least squares calculation. Second, under some conditions it can be shown that LASSO can recover the true sparseness of $\bbeta$ and deliver good prediction. However, the sufficient conditions to achieve this \citep{buhlmann2011statistics} are related to the model matrix $\bbX$, thus are difficult to be verified in practice. 

As argued in \cite{SCAD}, the original LASSO (\ref{eq:lasso}) leads to biased coefficient estimates. To address this shortcoming, several approaches such as adaptive LASSO \citep{adaptive} and SCAD \citep{SCAD} are proposed. It has been shown that their estimates possess the so-called oracle properties: (1) identifies the true set of predictors asymptotically; (2) has the optimal convergence rate. In our simulation study and real data application, we choose adaptive LASSO as a representative of $l_1$ regularization approaches, with objective function shown as follows:
\begin{equation}
\min_{\bbeta}\;\lbk\norm{\by - \beta_0\bone - \bbX\bbeta}^2_2+\lambda\sum_{i=1}^pw_i|\beta_i|\rbk.
\label{eq:adalasso}
\end{equation}
Here, $w_i$ ($i=1,\dots,p$) is the penalty weight of $\beta_i$ that can be derived from $\sqrt{T}$-consistent estimator such as LASSO estimator ($w_i=1/|\hat{\beta}^{LASSO}_i|$) or ridge estimator ($w_i=1/|\hat{\beta}^{ridge}_i|$).

In spite of the good properties mentioned above, $l_1$ regularization is a relaxation of VS problem, thus do not provide provably optimal solution to (\ref{eq:vsopt}). In addition, as shown in our simulation studies, for very sparse cases, $l_1$ regularization tends to introduce a lot of spurious variables. As a consequence, it may be inferior to other types of methods in terms of identifying the important predictors. 

\subsection{Gradient Decent Algorithms with Sparsification}
In the fields of computer science and signal processing, a group of algorithms, referred to gradient decent with sparsification (GDS) algorithms, are developed to directly provide a good solution to the $l_0$-constraint optimization (\ref{eq:vsopt}). As the name suggests, these methods are extensions of gradient descent algorithms which impose sparsity on the parameter estimate. These algorithms include but not limited to iterative hard thresholding \citep{BLUMENSATH2009}, compressive sampling matching pursuit \citep{CoSaMP2009}, subspace pursuit \citep{subspace2009}, hard thresholding pursuit \citep{HTP2011} and orthogonal matching pursuit with replacement \citep{orthogonal2011}. The analysis of their theoretical properties in high dimensional regression setting can be found in 
\cite{IHTtheory2014}, \cite{Bertsimas2016} and references therein. In general, this group of methods aim to minimize a loss function $f(\bbeta)$ subject to the $l_0$ constraint:
\begin{equation}
\min_{\bbeta}f(\bbeta) \quad\text{subject to} \quad \norm{\bbeta}_0\leq k.
\label{PGDopt}
\end{equation}
In our case, $f(\bbeta)$ equals to $\norm{\by - \beta_0\bone - \bbX\bbeta}^2_2$, the SSE of linear regression. Without considering the $l_0$ constraint, gradient descend algorithm minimizes $f(\bbeta)$ iteratively by updating $\bbeta$ as:
\begin{equation}
\bbeta^{(r+1)} = \bbeta^{(r)} - \eta\nabla f(\bbeta^{(r)}),
\label{eq:GD}
\end{equation}
where superscript $(r)$ stands for the $r$th iteration, $\nabla f(\bbeta^{(r)})=2\bbX'\bbX\bbeta^{(r)}-2\bbX'\by$ is the gradient of $f(\bbeta)$ at $\bbeta^{(r)}$, and $\eta$ is the step size. Since a $k$-sparse vector is desired, GDS-type algorithms modify the updating equation (\ref{eq:GD}) in some ways so that $\bbeta^{(r+1)}$ becomes a $k$-sparse vector. In the following, we will introduce four GDS-type algorithms.

Iterative hard thresholding (IHT), also known as projected gradient descent, modifies the updating equation by adding a hard thresholding operator $H_k(.)$ to it:
\begin{equation}
\text{IHT:}\quad \bbeta^{(r+1)} = H_k\lsk \bbeta^{(r)} - \eta\nabla f(\bbeta^{(r)})\rsk.
\end{equation}
For any vector $\bv$, the hard thresholding operator $H_k(\bv)$ keeps the $k$ largest (in magnitude) elements of $\bv$ and set the rests to zero. This operator is the simplest way to obtain a $k-$sparse vector. It has been also proven that for any arbitrary vector $\bv$, $H_k(\bv)$ is the closest $k$-sparse vector to it in $l_2$ distance.

IHT directly converts $\bbeta^{(r)} - \eta\nabla f(\bbeta^{(r)})$ into a sparse vector, while some other methods,  such as compressive sampling pursuit (CoSaMP) and subspace pursuit (SP), chase a good support of $\bbeta$ based on the gradient descent and then finds the best fit within this support. Specifically, the CoSaMP algorithm iterates the following three-step scheme:
\begin{align}
\text{CoSaMP:}\quad &U^{(r+1)} = \supp\lsk\bbeta^{(r)}\rsk \cup \lbk\text{indices of 2$k$ largest elements of }\nabla f(\bbeta^{(r)})\rbk, \\
&\tilde{\bbeta}^{(r+1)} = \argmin_{\supp\lsk\bbeta\rsk\subseteq U^{(r+1)}}\norm{\by - \beta_0\bone - \bbX\bbeta}^2_2, \\
&\bbeta^{(r+1)} = H_k\lsk \tilde{\bbeta}^{(r+1)}\rsk.
\end{align}
SP algorithm is similar to CoSaMP, except that it replaces $2k$ with $k$ in the first step of CoSaMP and modifies the third step:
\begin{align}
\text{SP:}\quad &U^{(r+1)} = \supp\lsk\bbeta^{(r)}\rsk \cup \lbk\text{indices of $k$ largest elements of }\nabla f(\bbeta^{(r)})\rbk, \\
&\tilde{\bbeta}^{(r+1)} = \argmin_{\supp\lsk\bbeta\rsk\subseteq U^{(r+1)}}\norm{\by - \beta_0\bone - \bbX\bbeta}^2_2, \\
&\bbeta^{(r+1)} = \argmin_{\supp\lsk\bbeta\rsk\subseteq V^{(r+1)}}\norm{\by - \beta_0\bone - \bbX\bbeta}^2_2,\; V^{(r+1)}=\lbk \text{$k$ largest elements of } \tilde{\bbeta}^{(r+1)} \rbk.
\end{align}
Unlike IHT, CoSaMP and SP have one and two OLS calculations respectively. Since OLS is the most time consuming part in each iteration, these two methods have more computational costs than IHT within one iteration. However, on the other hand, OLS offers the best fit within the proposed support, which may make these two methods converge with less iterations. 

\cite{HTP2011} combined the hard thresholding operator in IHT and the idea of pursuing a good support of $\bbeta$ (CoSaMP and SP) into the following hard thresholding pursuit (HTP) algorithm: 
\begin{align}
\text{HTP:}\quad &\tilde{\bbeta}^{(r+1)} = H_k\lsk\bbeta^{(r)} - \eta\nabla f(\bbeta^{(r)})\rsk,\\
&\bbeta^{(r+1)} = \argmin_{\supp\lsk\bbeta\rsk\subseteq \supp\lsk\tilde{\bbeta}^{(r+1)}\rsk}\norm{\by - \beta_0\bone - \bbX\bbeta}^2_2.
\end{align}
The difference between HTP and CoSaMP/SP lies on their ways of proposing the support of $\bbeta$. HTP uses the support of IHT result while CoSaMP and SP derives the support by the result of previous iteration $\bbeta^{(r)}$ and the corresponding gradient $f(\bbeta^{(r)})$. 

In general, GDS-style algorithms are not limited to the aforementioned four algorithms, and there is no guarantee that one algorithm outperforms the others. In this paper, we select IHT and HTP as representatives of this group of methods. The initial input $\bbeta^{(0)}$ should be a $k$-sparse vector, typically $\bbeta^{(0)}=\bzero$. The iteration is stopped when the difference between $\bbeta^{(r+1)}$ and $\bbeta^{(r)}$ is small enough or the maximum number of iterations is reached. 

\subsection{Sequential Monte Carlo \citep{Duan2019}}
Optimization (\ref{eq:vsopt}) is a combinatory optimization problem whose search space is discrete. Some meta-heuristic algorithms, such as simulated annealing \citep{SA1983,Cerny1985} and genetic algorithms \citep{goldberg2006genetic}, are dedicated to solve such combinatory optimization, and become appealing to variable selection \citep{chatterjee1996genetic,brusco2014,brooks2003}. 
Unlike greedy algorithms that always reject worse solutions,  meta-heuristic algorithms accept worse solution with a probability to yield a more extensive research. 
In this subsection, we introduce one meta-heuristic algorithm proposed by \cite{Duan2019}, called sequential Monte Carlo (SMC). This algorithm incorporates the idea of simulated annealing and Monte Carlo methods to considerably extend the searching space. Even the key idea involves Monte Carlo, SMC algorithm is different from the Bayesian variable selection methods that assume hierarchical distribution for the data and variables are selected by imposing some spike-and-slab priors on the model parameters. Indeed, SMC approach is not a Bayesian method. It does not require any distributional assumption and solves the $l_0$ constraint optimization (\ref{eq:vsopt}) in its original form.

For any nonempty set of indices $U$, let $\bbX_U=[\bx_i]_{i\in U}$ be the sub-matrix of $\bbX$ whose columns belong to set $U$, and let $\bbeta_U=[\beta_i]_{i\in U}$ be the corresponding sub-vector of coefficients. Then the $l_0$ constraint optimization (\ref{eq:vsopt}) is rewritten as a minimization problem:
\begin{equation}
\min_{|U|=k}\norm{\by-\hat{\beta}_0\bone-\bbX_U\hbbeta_U}^2_2, 
\label{eq:SMCobj}
\end{equation}
where $\hat{\beta}_0$ and $\hbbeta_U$ are the OLS estimates corresponding to $\bbX_U$. Let $\bbU_k$ be the collection of all the permutations of $k$ indices, i.e. $\bbU_k=\{U\subseteq\{1,\cdots,p\}: |U|=k\}$. \cite{Duan2019} assigned a discrete distribution function $f(U)$ on $\bbU_k$:
\begin{equation}
f(U)\propto \exp\lbk -\norm{\by-\hat{\beta}_0\bone-\bbX_U\hbbeta_U}^2_2 \rbk.
\label{eq:SMCtargetf}
\end{equation} 
Then finding the global optimum of (\ref{eq:SMCobj}) is done through generating a representative sample from this distribution. Obviously, among all permutations in $\bbU_k$, the optimal one corresponds to the peak of this distribution thus is more likely to be generated. Distribution $f$ is defined over permutations instead of combinations because permutations are easier to sample. Since this distribution has no tractable analytical solution and is multimodal, it is hard to directly construct a suitable proposal distribution for it. SMC applies the idea of distribution tempering that starts from an easy-to-sample distribution $f_0$, moves ``smoothly" to the complex target distribution $f$ by composing a sequence of artificial intermediate distributions:
\begin{equation}
f_0,f_1,\dots,f_J,\quad f_j(U)\propto f^{\gamma_j}(U)f^{1-\gamma_j}_0(U).
\end{equation}
The distribution sequence $\{f_j\}_{j=1}^J$ is called ``distribution-tempering bridge" and sequence $\{\gamma_j\}$ satisfies $0=\gamma_0<\gamma_1<\cdots<\gamma_J=1$. Clearly, $\gamma_0=0$ corresponds to the initial distribution $f_0$, and $\gamma_n=1$ corresponds to target distribution $f$. For each round $j$, the choice of $\gamma_j$ is self adapted by the algorithm, which makes the difference between $f_{j-1}$ and $f_j$ small enough so that the former distribution, $f_{j-1}$, can be a good proposal for the latter distribution, $f_j$. 

The initial distribution, $f_0$, takes into consideration the prediction power of each individual predictor. Let $R^2_i$ be the $R^2$ of a single linear regression with the $i$-th predictor ($i=1,\dots,p$). The initial distribution, $f_0$, is the random sampling without replacement based on inclusion probability $q_i=R^2_i/\lsk\sum_{i=1}^pR^2_i\rsk$:
\begin{equation}
f_0(\{i_1,i_2,\cdots,i_k\})=q_{i_1}\frac{q_{i_2}}{1-q_{i_1}}\frac{q_{i_3}}{1-q_{i_1}-q_{i_2}}\cdots\frac{q_{i_k}}{1-q_{i_1}-\cdots-q_{i_{k-1}}},\quad \{i_1,i_2,\cdots,i_k\}\in\bbU_k.
\label{eq:f0}
\end{equation}

After obtaining the initial sample, denoted by $\Omega_0=\{U_{0,m}\}_{m=1}^M$ with $M$ being the sample size, 
we can generate a representative sample for the next distribution $f_1$ by following the three-step scheme:
\begin{itemize}
	\item[] {\bf Reweighting:} First, an importance weight is assigned to each $U_{0,m}$, which is the ratio between the probability of $U_{0,m}$ under distribution $f_1$ and that under $f_0$, i.e. $w_{1,m}\propto f_1(U_{0,m})/f_0(U_{0,m})$. The parameter $\gamma_1$ in $f_1$ is determined to guarantee the effective sample size implied by these weights not smaller than a threshold, say $M/2$.
	\item[] {\bf Resampling:} Use the importance weights to resample the $U_{0,m}$'s in $\Omega_0$ which will result a new sample satisfying the distribution $f_1$.
	\item[] {\bf Support boosting:} After resampling, some $U_{0,m}$'s are duplicated to reflect their relatively high importance weights while some are excluded due to their low weights. Thus the empirical support (distinct $U$'s in the sample) is shrunk. Boosting the empirical support is accomplished by several Metropolis Hastings moves which reduce duplicates, enlarge the number of distinct members, and retain distribution $f_1$ for the sample at the meantime. After this step, we will get a representative sample for $f_1$, denoted as $\Omega_1=\{U_{1,m}\}_{m=1}^M$.
\end{itemize}
By repeating this three-step scheme until $\gamma$ reaches 1, we finally arrive at a representative sample for the target distribution $f$. 
At last, a $k$-fold duplication technique \citep{Duan2015} is carried out to enlarge the sample size $k$ times, and the best subset of predictors is the $U$ with the smallest SSE in the final sample. The detailed algorithm can be found in Appendix \ref{subsec:appSMC}, and readers can refer to \cite{Duan2019} for its theoretical properties. 

\section{Simulation Studies}
\label{chap4:simulation}

We present a variety of computational experiments to:  (1) evaluate different types of VS methods in terms of variable selection accuracy and out-of-sample prediction; (2) compare FA approach to VS approach under the framework of time series forecasting. These two goals are addressed in Section \ref{subchap4:simu1} and \ref{subchap4:simu2} respectively. 

\subsection{Evaluation of Different Types of VS Methods}
\label{subchap4:simu1}
We conduct three simulation studies to compare among different VS methods in terms of variable selection accuracy and out-of-sample prediction. For each group of VS methods, we select one or two representatives: (1) FS; (2) adaptive LASSO (adaLASSO); (3) IHT and HTP algorithms; (4) SMC algorithm. In our simulation studies, we investigate the $p>T$ case, thus BE is not applicable and excluded. 
In order to examine their performance, we consider five criteria:
\begin{itemize}
	\item precision = True Positive/(True Positive + False Positive);
	\item recall = True Positive/(True Positive + False Negative);
	\item dice coefficient (DC) = 2 True Positive/(2 True Positive + False Positive + False Negative);
	\item mean squared prediction error (MSPE) on test set (sample size=100).
	\item time cost in minutes for one simulation.
\end{itemize}
Here, ``True Positive" stands for the number of predictors that are correctly identified, ``False Positive" is the number of predictors which are wrongly selected, and ``False Negative" counts the number of predictors belonging to the true model but missed by the VS method. In this way, recall is the proportion of predictors in the true model that correctly identified, while precision is the proportion of the selected predictors that are truly significant. These two criteria quantify different aspects of the variable selection accuracy, and higher value indicates higher accuracy. However, these is a trade-off between precision and recall: as more predictors are selected, precision tends to increase while recall would decrease. The criteria, DC, can balance this trade-off and serves as an overall measurement of variable selection accuracy. Since this paper aims to apply VS methods on economic forecasting,  we also include the MSPE as the criterion to evaluate out-of-sample prediction. 

\subsubsection{Simulation Settings}
Simulation is carried out under the linear regression framework. The true model is:
\begin{equation}
    y_t=\sum_{j=1}^K\beta_jx_{jt} + \epsilon_t, \quad  \epsilon_t \overset{\text{i.i.d}}{\sim} N(0, \sigma^2), \quad t=1,\dots,T.
\end{equation}
The regressors are selected from $p$ potential predictors with $p\gg K$. All the potential predictors are generated from normal distribution with mean zero and variance one. Note that in this subsection we do not consider temporal dependence thus the observations are independent across time. In total, we construct three settings to evaluate VS methods under different scenarios. The first one adopts the simulation setting of \cite{Duan2019}, which presents moderate high dimensional case ($p=900$ and $T=200$). The other two settings are cases of ultra-high dimension ($p=2000$, $T=50,\;100,\;200$), with setting 2 imposing independent structure and setting 3 imposing correlated structure among predictors.
\begin{itemize}
	\item[]{\bf Setting 1:} 900 potential predictors are equally divided into three groups with 300 in each. The correlation within each group are 0.1, 0.4 and 0.8 respectively, and predictors across different groups are independent. Within each group, four predictors are included in the true model with coefficients 0.1, 0.4, 0.7 and 1. In total there are $K=12$ relevant predictors. To ascertain the impact of signal strength in variable selection, two levels of theoretical $R^2$ ( 0.8 and 0.5) are considered, which determine the values of $\sigma^2$.
	\item[]{\bf Setting 2:} 2000 potential predictors  are  generated independently from $N(0,1)$. The true model only includes five of them ($K=5$) and all of their coefficients are one. The theoretical $R^2$ is set to be 0.8. Three different sample sizes are considered ($T=$50, 100 and 200) to assess its influence on model performance.
	\item[]{\bf Setting 3:} 2000 potential predictors are divided into four groups with 500 in each. The correlation within each group are 0.1, 0.4, 0.7 and 0.9 respectively, and predictors across different groups are independent. The true model includes two predictors from each group with coefficient 1, total $K=8$ predictors in the model. The theoretical $R^2$ is set to 0.8. Similar to setting 2, we also consider three sample sizes: $T=$50, 100 and 200.
\end{itemize}
All simulations are conducted with 100 repetitions. For FS, IHT, HTP and SMC, the tuning parameter is the the subset size $k$, while for adaLASSO, the tuning parameter is the penalty factor $\lambda$. Both of them are selected by 5-fold cross validation.

\subsubsection{Simulation Results}
\begin{table}[H]
	\centering
	\caption{Five quantiles of the number of selected predictors by each VS method and the mean of in-sample $R^2$ among 100 repetitions for setting 1.}
	\begin{tabular}{ccccccccccc}
		\hline
		& \multicolumn{5}{c}{{\bf Theoretical} $\boldsymbol{R^2=0.8}$} &  \multicolumn{5}{c}{{\bf Theoretical} $\boldsymbol{R^2=0.5}$}\\\cmidrule(r){2-6}\cmidrule(r){7-11}
		& FS & SMC & adaLASSO & IHT & HTP & FS & SMC & adaLASSO & IHT & HTP \\ 
		Min. & 5 & 5 & 16 & 6 & 5 & 5 & 5 & 13 & 5 & 5 \\ 
		1st Qu. & 7 & 7 & 32 & 9 & 8 & 7 & 7 & 26 & 7 & 7 \\ 
		Median & 8 & 7 & 38 & 10 & 8 & 7 & 7 & 34 & 7 & 7 \\ 
		3rd Qu. & 9 & 8 & 46 & 11 & 10 & 7 & 8 & 42 & 8 & 7 \\ 
		Max. & 13 & 12 & 63 & 17 & 14 & 11 & 12 & 83 & 13 & 11 \\ 
		\hline
		$R^2$ & 0.80 & 0.80 & 0.85 & 0.76 & 0.78  & 0.55 & 0.57 & 0.64 & 0.51 & 0.52 \\ 
		\hline
	\end{tabular}
	\label{table:kset1}
\end{table}

Table \ref{table:kset1} lists the five quantiles of the number of selected predictors by each VS method in setting 1, and the mean of $R^2$ in 100 repetitions. Recall that the true model contains 12 predictors by design. Clearly adaLASSO tends to pick too many predictors, with median of 38 and 34 under the two scenarios of theoretical $R^2$ respectively. Moreover, the range is much wider than those from other methods, reflecting its low stability in selecting predictors. This is not surprising since adaLASSO fails to distinguish between a zero and a nonzero coefficient when the corresponding predictors are correlated. 

\begin{table}[H]
	\centering
	\caption{Hit ratio of each predictor in the true model in setting 1. For each predictor, ``corr" is the within-group correlation, ``coef" is the coefficient, and hit ratio is the proportion of the 100 repetition that successfully identified the predictor. }
	\begin{tabular}{cccccccccccc}
		\hline
		corr & coef & \multicolumn{5}{c}{{\bf Theoretical} $\boldsymbol{R^2=0.8}$} &  \multicolumn{5}{c}{{\bf Theoretical} $\boldsymbol{R^2=0.5}$}\\\cmidrule(r){3-7}\cmidrule(r){8-12}
		& & FS & SMC & adaLASSO & IHT & HTP & FS & SMC & adaLASSO & IHT & HTP \\ 
		\multirow{4}{*}{0.1} & 0.1 & 0.02 & 0.03 & 0.17 & 0.01 & 0.02 & 0.01 & 0.02 & 0.09 & 0.01 & 0.01 \\ 
		& 0.4 & 0.44 & 0.38 & 0.82 & 0.15 & 0.42 & 0.10 & 0.10 & 0.34 & 0.02 & 0.08 \\ 
		& 0.7 & 0.95 & 0.98 & 1.00 & 0.68 & 0.91 & 0.41 & 0.42 & 0.79 & 0.25 & 0.36 \\ 
		& 1.0 & 1.00 & 1.00 & 1.00 & 0.97 & 0.98 & 0.81 & 0.84 & 0.98 & 0.71 & 0.76 \\ 
		\hline
		\multirow{4}{*}{0.4} & 0.1 & 0.01 & 0.01 & 0.14 & 0.01 & 0.03 & 0.04 & 0.03 & 0.14 & 0.03 & 0.04 \\ 
		& 0.4 & 0.32 & 0.30 & 0.74 & 0.12 & 0.23 & 0.13 & 0.18 & 0.35 & 0.10 & 0.10 \\ 
		& 0.7 & 0.89 & 0.93 & 0.99 & 0.60 & 0.70 & 0.31 & 0.33 & 0.65 & 0.23 & 0.30 \\ 
		& 1.0 & 1.00 & 1.00 & 1.00 & 0.96 & 0.98 & 0.72 & 0.73 & 0.92 & 0.63 & 0.69 \\ 
		\hline
		\multirow{4}{*}{0.8} & 0.1 & 0.02 & 0.02 & 0.08 & 0.03 & 0.01 & 0.01 & 0.00 & 0.03 & 0.01 & 0.00 \\ 
		& 0.4 & 0.06 & 0.04 & 0.27 & 0.10 & 0.02 & 0.05 & 0.07 & 0.17 & 0.09 & 0.06 \\ 
		& 0.7 & 0.35 & 0.40 & 0.68 & 0.26 & 0.22 & 0.08 & 0.11 & 0.28 & 0.17 & 0.07 \\ 
		& 1.0 & 0.79 & 0.88 & 0.97 & 0.51 & 0.45 & 0.15 & 0.22 & 0.47 & 0.24 & 0.16 \\ 
		\hline
	\end{tabular}
	\label{table:hitset1}
\end{table}

In contrast to adaLASSO, the other four methods are more conservative and tends to select less predictors than the true number 12. For example, FS yields a median of 8 predictors and covers the range from 5 to 13. Such under-selection is actually expected, because cross validation is a conservative way to find the correct number of predictors by avoiding  over-fitting. In addition, the under-selection does not ruin the model fitting performance that is measured by in-sample $R^2$. The $R^2$ obtained by FS, SMC, IHT and HTP are close to the theoretical $R^2$. To better understand the reason of under-selection, we calculate the hit ratio (the proportion of successfully identifying the predictor in 100 repetitions) of each predictor in the true model, as shown in Table \ref{table:hitset1}. It appears that the predictors with small magnitudes of coefficient (0.1 and 0.4) would be excluded from the model due to their relatively low prediction power, especially when they are highly correlated with other predictors and the sample size ($T=200$) is not very large. It is also natural to see that the hit ratio is higher when the signal strength is higher (theoretical $R^2$=0.8) and the predictors are less correlated. 

\begin{table}[H]
	\centering
	\caption{Simulation results in Setting 1. the column ``true" lists the results of the true model. For each criteria, the mean of 100 repetitions is presented, followed by the standard error in the parenthesis.}
	\begin{tabular}{ccccccc}
		\hline
		& true & FS &  SMC & adaLASSO & IHT & HTP \\ 
		\hline
		\multicolumn{7}{l}{{\bf Theoretical} $\boldsymbol{R^2=0.8}$} \\
		\multirow{2}{*}{MSPE} & 2.43 & 3.07  & 2.97 & 2.99 & 3.44 & 3.55 \\ 
		& (0.04) & (0.05) & (0.05) & (0.05) & (0.08) & (0.09) \\ 
		\multirow{2}{*}{DC} & -- & 0.59 & 0.61 & 0.32 & 0.40 & 0.48 \\ 
		& -- & (0.01) & (0.01) & (0.01) & (0.01) & (0.01) \\ 
		\multirow{2}{*}{Precision} & --& 0.74 & 0.79 & 0.22 & 0.45 & 0.57 \\ 
		& -- & (0.01) & (0.01) & (0.01) & (0.02) & (0.01) \\ 
		\multirow{2}{*}{Recall} & -- & 0.49 & 0.50 & 0.65 & 0.37 & 0.41 \\ 
		& -- & (0.01) & (0.01) & (0.01) & (0.01) & (0.01) \\ 
		\multirow{2}{*}{time} & -- & 0.04 & 168.25 & 0.01 & 0.29 & 0.75 \\ 
		& -- & (0.00) & (0.76) & (0.00) & (0.00) & (0.00) \\ 
		\hline
		\multicolumn{7}{l}{{\bf Theoretical} $\boldsymbol{R^2=0.5}$} \\
		\multirow{2}{*}{MSPE} & 9.59 & 12.58 & 12.79 & 11.33 & 11.96 & 12.66 \\ 
		& (0.16) & (0.24) & (0.24) & (0.17) & (0.20) & (0.24) \\ 
		\multirow{2}{*}{DC}	 & -- & 0.30 & 0.31 & 0.23 & 0.25 & 0.27 \\ 
		& -- & (0.01) & (0.01) & (0.01) & (0.01) & (0.01) \\ 
		\multirow{2}{*}{precision} & --& 0.41 & 0.41 & 0.16 & 0.32 & 0.37 \\ 
		& -- & (0.02) & (0.02) & (0.01) & (0.02) & (0.02) \\
		\multirow{2}{*}{recall} & -- & 0.24 & 0.26 & 0.44 & 0.21 & 0.22 \\ 
		& -- & (0.01) & (0.01) & (0.01) & (0.01) & (0.01) \\
		\multirow{2}{*}{time}	& -- & 0.04 & 162.68 & 0.02 & 0.29 & 0.75 \\ 
		& -- & (0.00) & 0.63 & (0.00) & (0.00) & (0.00) \\ 
		\hline
	\end{tabular}
	\label{table:set1}
\end{table}

Table \ref{table:set1} summarizes the results of the five criteria in setting 1. When the signal is strong (theoretical $R^2=0.8$), SMC, adaLASSO and FS are the best group in terms of prediction performance, which are significantly superior than IHT and HTP. As for the variable selection accuracy, SMC and FS outperforms the rests. Since adaLASSO suffers from excessive over-selection, its precision and DC are the lowest while its recall is the highest. 
HTP is a little better than IHT but still worse than SMC and FS. As for time cost, SMC is the most expensive algorithm, which takes about three hours to finish one simulation while others cost less than one minute. FS and adaLASSO are the most time-efficient, which only take two seconds or less to finish one simulation.  In summary, the most sophisticated algorithm, SMC, is the best in terms of both variable selection accuracy and out-of-sample prediction but extremely time consuming. FS, the classical greedy algorithm, obtains very similar performance as SMC while it is much more computationally efficient. If we only focus on prediction, adaLASSO is one of the best choices given its great efficiency and promising prediction results.

In the scenario of theoretical $R^2=0.5$, the differences in prediction among all five methods are quite small, while adaLASSO outperforms the rests a little bit. As for variable selection accuracy, adaLASSO is still inferior in terms of DC and precision. SMC and FS are very close to each other and significantly better than IHT and HTP. This scenario asserts that, when the signal is not strong enough, adaLASSO is preferred for prediction while FS and SMC can provide more accurate variable selection results.

In setting 2 and 3, we only report the results of MSPE and DC in Table \ref{table:set2} and \ref{table:set3}, and leave other criteria in Appendix \ref{sec:appA} (Table \ref{table:set2precisionrecall} and \ref{table:set3precisionrecall}). it can be found that all the VS methods work better in the independent case (setting 2) than in dependent case (setting 3). In addition, as sample size $T$ increases, their results become closer to that of the true model. In setting 2, $T=50$ is too small to successfully recover the true sparsity or obtain a satisfying prediction. When sample size increases to 100, SMC and FS have DC as 0.96 and 0.98 respectively which means they almost completely recover the true model, and their prediction are also
quite close to the true model. While adaLASSO, IHT and HTP are still falling behind. When sample size increases to 200, except for adaLASSO which still over-selects predictors, all other VS methods are almost the same as the true model. In setting 3, since the potential predictors are correlated, VS methods can not provide satisfying performance until sample size increases to 200. Under this sample size, FS and SMC have the smallest prediction error and provide the most accurate models, followed by adaLASSO whose prediction is close to SMC and FS but performs worse in variable selection. In both settings, when the sample size is not large enough ($T=50$ in setting 2 and $T=50$ or 100 in setting 3), adaLASSO has the best prediction, followed by SMC, and SMC is superior than others in terms of identifying the true predictors.

\begin{table}[H]
	\centering
	\caption{Out-of-sample MSPE and in-sample DC in setting 2. The column ``true" lists the results of the true model. The mean of 100 repetitions is presented, followed by the standard error in the parenthesis.}
	\begin{tabular}{ccccccccccc}
		\hline
		\multicolumn{6}{c}{{\bf MSPE}} & \multicolumn{5}{c}{{\bf DC}} \\\cmidrule(r){1-6}\cmidrule(r){7-11}
		true & FS &  SMC & adaLASSO & IHT & HTP & FS & SMC & adaLASSO & IHT & HTP \\  
		\multicolumn{11}{l}{$\boldsymbol{T=50}$} \\
		1.35 & 5.96  & 5.43 & 4.78 & 5.69 & 5.84 & 0.31 & 0.37 & 0.19 & 0.28 & 0.26 \\ 
		(0.03) & (0.20)  & (0.22) & (0.14) & (0.13) & (0.14) & (0.02) & (0.03) & (0.01) & (0.02) & (0.02) \\  
		\multicolumn{11}{l}{$\boldsymbol{T=100}$} \\
		1.32 & 1.41  & 1.36 & 2.26 & 2.87 & 2.56 & 0.96 & 0.98 & 0.19 & 0.70 & 0.76 \\ 
		(0.02) & (0.03) & (0.03) & (0.06) & (0.11) & (0.12) & (0.01) & (0.00) & (0.01) & (0.02) & (0.02) \\ 
		\multicolumn{11}{l}{$\boldsymbol{T=200}$} \\
		1.28 & 1.29 & 1.29 & 1.60 & 1.33 & 1.28  & 0.99 & 0.99 & 0.23 & 0.98 & 0.99 \\  
		(0.02) & (0.02) & (0.02) & (0.03) & (0.02) & (0.02) & (0.00) & (0.00) & (0.02) & (0.00) & (0.00) \\  
		\hline
	\end{tabular}
	\label{table:set2}
\end{table}

\begin{table}[H]
	\centering
	\caption{Out-of-sample MSPE and in-sample DC in setting 3. The column ``true" lists the results of the true model. The mean of 100 repetitions is presented, followed by the standard error in the parenthesis.}
	\begin{tabular}{ccccccccccc}
		\hline
		\multicolumn{6}{c}{{\bf MSPE}} & \multicolumn{5}{c}{{\bf DC}} \\\cmidrule(r){1-6}\cmidrule(r){7-11}
		true & FS &  SMC & adaLASSO & IHT & HTP & FS & SMC & adaLASSO & IHT & HTP \\  
		
		\multicolumn{11}{l}{{$\boldsymbol{T=50}$}} \\
		3.67 & 11.93  & 11.31 & 7.97 & 11.78 & 12.16 & 0.06 & 0.08 & 0.13 & 0.04 & 0.04 \\  
		(0.07) & (0.31) & (0.28) & (0.17) & (0.28) & (0.28) & (0.01) & (0.01) & (0.01) & (0.01) & (0.01) \\  
		\multicolumn{11}{l}{{$\boldsymbol{T=100}$}} \\
		3.34 & 7.46 & 6.78 & 5.88 & 9.39 & 9.21 & 0.33 & 0.45 & 0.19 & 0.18 & 0.16 \\ 
		(0.06) & (0.21) & (0.23) & (0.13) & (0.27) & (0.25) & 0.02 & (0.02) & (0.01) & (0.02) & (0.02) \\  
		\multicolumn{11}{l}{{$\boldsymbol{T=200}$}} \\
		3.20 & 4.07 & 4.01 & 4.29 & 7.82 & 6.65 & 0.69 & 0.74 & 0.24 & 0.29 & 0.46 \\  
		(0.05) & (0.08) & (0.09) & (0.08) & (0.32) & (0.34) & (0.01) & (0.01) & (0.01) & (0.02) & (0.02) \\ 
		\hline
	\end{tabular}
	\label{table:set3}
\end{table}

We summarize our findings in these three simulation studies as follows. (1) When the signal is strong and the sample size is large enough, SMC and FS are the best among the five VS methods in terms of variable selection accuracy and prediction. However, SMC is very time consuming. The classical procedure FS is preferred if the time constraint is a concern. (2) adaLASSO is very good at prediction in all settings, especially when the signal is not strong or sample size is not large. However, it suffers from over-selection, and it is worse than others in terms of variable selection. (3)  In general, IHT is slightly better than HTP, but both are significantly worse than SMC and FS. (4) When the sample size is too small or signal is not strong, none of these method can provide good performance, and their difference are insignificant.

\subsection{Comparison between VS and FA}
\label{subchap4:simu2}
In this simulation study, we investigate FA versus VS approaches under the framework of time series forecasting. The dimension of predictors and number of observations mimics the scales of our real data in Section \ref{chap4:application}. The simulation setting is described as follows. 

We generate four groups of predictors with group id as $i=1$, 2, 3 and 4. In each group, a series of latent factor $\{f_{it}\}$ is generated based on a AR(1) process, then 30 predictors are generated from this latent factor: 
\begin{align}
&f_{it} = \phi_i f_{i,t-1} + e_{it}, \quad e_{it}\overset{\text{i.i.d}}{\sim} N(0,\delta_{i1}^2), \quad i=1,\cdots,4, \\
&x_{ijt} = f_{it} + e_{ijt}, \quad e_{ijt}\overset{\text{i.i.d}}{\sim} N(0,\delta_{i2}^2),\quad j=1,\dots,30, \quad t=1,\dots,T.
\end{align}
Clearly the series $\{x_{ijt}\}$ also follows a VAR(1) process with parameter $\phi_i$. Within one group, the temporal dependence is captured by the AR parameter $\phi_i$, while the cross-sectional correlation, cor$(x_{ijt},x_{ij't})$, is controlled by the relative magnitude between var($f_{it}$) and $\delta^2_{i2}$. By setting the values of $(\phi_i,\delta^2_{i1},\delta^2_{i2})$ as shown in Table {\ref{table:set4}}, the four groups are constructed to have different temporal and cross-sectional dependence. Specifically, both temporal and cross-sectional dependence are strong in group 1. While for group 2, only the temporal dependence is strong but the cross-sectional dependence is weak. Group 3 has the opposite pattern to group 2, and in group 4, both temporal and cross-sectional dependence are weak. 
\begin{table}[H]
	\centering
	\caption{Settings of predictors}
	\begin{tabular}{cccccc}
		\hline
		group $i$ & $\phi_i$ &  $\delta^2_{i1}$ & $\delta^2_{i1}$ & var($x_{ijt}$) & cor$(x_{ijt},x_{ij't})$ \\  
		\hline
		1 & 0.7 & 0.357 & 0.3 & 1 & 0.7 \\
		2 & 0.7 & 0.153 & 0.7 & 1 & 0.3 \\
		3 & 0.3 & 0.637 & 0.3 & 1 & 0.7 \\
		4 & 0.3 & 0.273 & 0.7 & 1 & 0.3 \\
		\hline
	\end{tabular}
	\label{table:set4}
\end{table}

For the response variable, we consider two generating mechanisms:
\begin{eqnarray}
&\text{Setting 1:}& y_t = \sum_{i=1}^4(0.8x_{i1t}+0.4x_{i2t}) + \sum_{i=1}^4(0.8x_{i1,t-1}+0.4x_{i2,t-1})  + \epsilon_{t}, \\
\label{eq:yger1}
&\text{Setting 2:}& y_t = \sum_{i=1}^4 0.8f_{it}+\sum_{i=1}^40.4 f_{i,t-1} + \varepsilon_{t} .
\label{eq:yger2}
\end{eqnarray}
Setting1 indicates that $y_t$ is a sparse linear function of the predictors and their lagged values, which favors VS approach over FA. In contrast, setting 2 implies that $y_t$ is directly generated from the latent factors, thus FA is the correct approach while VS approach miss-specifies the model. In both settings, the theoretical $R^2$ is set to be 0.8. 

In each setting, we generate 300 observations ($T=300$) and preserve the last 50 observations as test set to make prediction. We select FS, SMC, adaLASSO and IHT as representatives of VS approach and compare them with FA. For both FA and VS approaches, we do not include the lagged value of $y_t$ as predictor. Thus the models for FA and VS approaches are:
\begin{align}
&\text{FA approach: } \quad y_t = \sum_{l=0}^{m-1} \bgamma_l'\bbf_{t-l} +  \varepsilon_{t}, \quad \bZ_t = \bLambda \bF_t + \be_t,\\
&\text{VS approach: } \quad y_t = \bX_t'\bbeta + \epsilon_{t}.
\end{align}
To predict $y_t$, the vector $\bX_t$ in VS approach consists of $x_{ijt}$'s and their historical values up to lag 5, i.e. $\bX_t=\{x_{11t},x_{11,t-1},\dots,x_{11,t-5},\dots,x_{4,30,t},x_{4,30,t-1},\dots,x_{4,30,t-5}\}$,  with total of 720 predictors. While the vector $\bZ_t$ in FA approach only contains $x_{ijt}$'s, i.e. $\bZ_t=\{x_{11t},\dots,x_{4,30,t}\}$. The tuning parameters in VS methods are selected by FCV with validation set of size 50. For FA approach, we consider both BIC and FCV to determine the number of factors $d$ and the order of lagged factors $m$. The corresponding results are labeled as FA\_BIC and FA\_FCV respectively.  

\begin{table}[H]
	\centering
	\caption{Results of FA and VS approaches}
	\begin{tabular}{cccccccc}
		\hline
		& true & FA\_BIC & FA\_FCV & FS & SMC & adaLASSO & IHT \\ 
		\hline
		\multicolumn{8}{l}{\bf Setting 1: $\boldsymbol{y_t}$ generated from predictors} \\
		\multirow{2}{*}{MSPE} & 3.08 & 6.52 & 6.76 & 4.25 & 4.16 & 3.96 & 5.76 \\ 
		& (0.07) & (0.14) & (0.15) & (0.11) & (0.10) & (0.08) & (0.23) \\ 
		\multirow{2}{*}{$R^2$} &  0.81 & 0.61 & 0.63 & 0.80 & 0.82 & 0.84 & 0.72 \\ 
		& (0.00) & (0.00) & (0.01) & (0.00) & (0.00) & (0.00) & (0.01) \\ 
		\multirow{2}{*}{DC} & -- & -- & -- & 0.67 & 0.71 & 0.50 & 0.50 \\ 
		& -- & -- & -- & (0.01)  & (0.01) & (0.01) & (0.01) \\ 
		\multirow{2}{*}{Precision} & -- & -- & -- & 0.78 & 0.77 & 0.36 & 0.55 \\ 
		& -- & -- & -- & (0.02) & (0.02) & (0.01) & (0.01) \\ 
		\multirow{2}{*}{Recall}	 & -- & -- & -- & 0.61 & 0.68 & 0.87 & 0.48 \\ 
		& -- & -- & -- & (0.01) & (0.01) & (0.01) & (0.02) \\
		\hline
		\multicolumn{8}{l}{\bf Setting 2: $\boldsymbol{y_t}$ generated from factors} \\
		\multirow{2}{*}{MSPE} & 0.57 & 0.68 & 0.71 & 1.20 & 1.14 & 0.82 & 1.27 \\
		& (0.01) & (0.02) & (0.02) & (0.03) & (0.03) & (0.02) & (0.03) \\ 
		\multirow{2}{*}{$R^2$} & 0.81 & 0.79 & 0.80 & 0.80 & 0.81 & 0.88 & 0.71 \\ 
		& (0.00) & (0.00) & (0.00) & (0.01) & (0.00) & (0.00) & (0.01) \\ 
		\hline
	\end{tabular}
	\label{table:simu4}
\end{table}

Tale \ref{table:simu4} shows the results of comparison between FA and VS approaches. Unsurprisingly, VS methods are uniformly better than FA in setting 1 and are inferior to FA in setting 2. For FA approach, there is little difference between the two criteria, BIC and FCV. While among the four VS methods, their comparisons are similar to those in the first simulation study: SMC and adaLASSO yield the most precise prediction followed by FS with no significant difference. As for variable selection accuracy, SMC and FS are the best while adaLASSO over-selects predictors. Based on this simulation study, we can conclude that, if only a handful of predictors are relevant to the response variable, VS methods are more advantageous than FA approach. However if there are many relevant predictors which possess a factor structure, FA approach is more suitable. 


\section{Real-Time Macroeconomic Forecasting}
\label{chap4:application}

In this section, we apply both FA and VS approaches on forecasting of several important macroeconomic indices. By simulating real-time forecasting, we compare their empirical performance in real application. Our target indices are employment (EMP), industrial production index (IP) and consumer price index-all urban consumers (CPI). The predictors contains 128 economic variables in the FRED-MD dataset \citep{FREDMD}, which is available online: \url{research.stlouisfed.org/econ/mccracken/fred-databases}.

\subsection{Implementation Details}
We follow the framework of economic forecasting stated in Section \ref{chap4:ecoforecast}. Let $y^h_{t+h}$ be the $h$-month ahead value of the variable to be forecasted and $y_t$ be the corresponding $t$-dated value, which are required to be stationary. Following \cite{SW2002b} and \cite{BaiNg2008}, we define $y^h_{t+h}$ and $y_t$ as follows:
\begin{align}
&y_{t+h}^h =(1200/h)ln(EMP_{t+h}/EMP_t), \quad  y_t = 1200 ln(EMP_t/EMP_{t-1}), \\
&y_{t+h}^h =(1200/h)ln(IP_{t+h}/IP_t), \quad  y_t = 1200 ln(IP_t/IP_{t-1}), \\
&y_{t+h}^h = (1200/h)ln(CPI_{t+h}/CPI_t)-1200ln(CPI_t/CPI_{t-1}), \quad y_t = 1200\bigtriangleup ln(CPI_t/CPI_{t-1}).
\end{align}
We carry out the transformation proposed by \cite{FREDMD} to make the 128 economic variables in FRED-MD dataset stationary. Then each variable is standardized to have mean zero and variance one. 

After the transformation and standardization, the 128 economic variables compose the input vector $\bZ_t$ of factor model (\ref{eq:factor}) in the FA approach. As for the VS methods, the potential predictors contain both $\bZ_t$ and its historical values up to 5 months, i.e. $\bX_t=(\bZ_t,\dots,\bZ_{t-5})$, with dimension 768. Since both FA and VS approaches are under the framework of linear regression, their parameter estimation is sensitive to outliers. According to \cite{SW2002a,SW2002b}, in a series $\{z_{it}\}$, an outlier is defined as an observation that deviates from the sample median by more than ten interquantile ranges. After identifying the outliers, FA approach replaces them with missing values (NA), and estimates the factors using EM algorithms to account for missing values. As for VS approach, we can either delete the series contaminated by outliers or impute the values by kalman filter. Therefore, we consider two strategies to account for outliers in the predictors:
\begin{itemize}
	\item[1.] Remove the series with outliers: For fair comparison, we also delete these series in both VS and FA approaches;
	\item[2.] For FA approach, replace outliers with missing values and implement the EM algorithm. For VS methods, impute the outliers using kalman filter in R package \texttt{imputeTS}.
\end{itemize}
For the target variable $y$, it does not contain any outlier, i.e. no value is beyond the ten interquantile ranges from the median. In our empirical results, generally FA and VS approaches perform slightly better under strategy 1. Thus we only report the results of strategy 1 in the main context, and put the results of strategy 2 in Table \ref{table:real2} in Appendix \ref{sec:appA}.

There are three tuning parameters in FA approach: number of factors $d$,  lags of $y_t$, $q$, and lags of the factors $m$. We choose $(d, p, m)$ from $\{ 0 \le d \le 5, 0 \le q \le 6, 1 \le m \le 6 \}$ by FCV, where $d=0$ means no factor is included in the forecasting equation (\ref{eq:faforecast}) and $p=0$ means no auto-regressive term is included. The tuning parameter for VS is the number of selected predictors, which is also determined by FCV. In addition, we include univariate AR as benchmark model in the comparison:
\begin{equation}
y_{t+h}^h = \alpha + \sum_{l=0}^{\tilde{p}-1} \alpha_l y_{t-l} + \varepsilon_{t+h},
\label{eq:ar}
\end{equation}
where the AR order $\tilde{p}$ is selected from $0\leq \tilde{p} \leq 6$ by FCV. 

The time span from 2015:1 to 2018:12 is reserved as test set to evaluate out-of-sample forecasting, with forecast horizon $h$ being 1, 3, 6 or 12 months. For both FA and VS approaches, the estimations and forecasts are conducted to simulate real-time forecasting using a rolling window scheme with window size 240. For example, if we want to forecast the $h$-month growth rate of EMP on 2015:1, The target variable is $y^h_{t+h}$ with $t\,$= 2015:1$\,-\,h$. Here 2015:1$\,-\,h$ stands for the month that is $h$ months before 2015:1. For example, if $h=3$, 2015:1$\,-\,h$ is 2014:10. To obtain its forecast, both FA and VS use data from 1995:1$\,-\,h$ to 2015:1$\,-\,h$ (240 months) to estimate model and select tuning parameters. The validation set for selecting tuning parameters is the sub-sample from 2011:1$\,-\,h$ to 2015:1$\,-\,h$ (48 months). When forecasting $y^h_{t+h}$ on 2015:2 ($t\,$= 2015:2$\,-\,h$), the tuning parameters are re-selected and the models are re-estimated using data from 1995:2$\,-\,h$ to 2015:2$\,-\,h$ (validation set is 2011:2$\,-\,h$ to 2015:2$\,-\,h$). In this way, we allow the parameter estimation and the optimal number of predictors and factors to change across $t$. For each pair of target variable and forecasting horizon $h$, a method produces 48 forecasts on the test set. We use MSPE of these 48 forecasts to evaluate the forecasting performance of this method. The ratio between MSPE of each method and that of the benchmark model (\ref{eq:ar}) is reported. Ratio less than one indicates the method has smaller forecasting error than the univariate AR.

\subsection{Empirical Results}

The MSPE ratios are presented in Table \ref{table:real}. For each forecast horizon $h$, we define the best group of results as those with MSPE ratios which do not exceeding 105\% of the smallest one. These best ratios are marked in bold. For each target variable, the predictors\footnote{We only report the predictors selected with at least 12 times in the 48 rolling-window forecasts. The predictors with frequency less than 12 are viewed as lacking of systematic prediction power.} selected by the best VS method and their number of occurrences in the 48 forecasts are reported in Table \ref{table:sltvar} in Appendix \ref{sec:appA}. 

For EMP, all the ratios are smaller than one, which manifests the usefulness of incorporating many predictors in forecasting. Among different approaches, SMC ranks the best for all the forecasting horizons. For $h=1$ and 3, the improvement of SMC over FA is not obvious, while for $h=6$ and 12, SMC reduces the MSPE of FA approach by 33\% and 52\% respectively. 

The success of SMC implies EMP can be forecasted by only a few predictors. This point is also demonstrated in Table \ref{table:sltvar}, which shows that only two to five predictors are selected for each horizon. Specifically,  $t$-dated EMP is selected in almost all 48 forecasts for all horizons, which implies the historical value has significant prediction power for EMP. For $h=6$ and 12, the linear forecasting model also includes two interest rates: 3-month treasury C minus FEDFUNDS and 3-month commercial paper minus FEDFUNDS. This leading effect of interest rate on EMP can be explained by economic theory. With low interest rate, consumers are more likely to consume now rather than wait for later. Low interest rate also drop the cost of borrowing to invest. Thus the increase in consumption and investment leads to higher demand for labor. As for one-month ahead forecast ($h=1$), except for the $t$-dated value of EMP, the model also includes M2 money stock, real personal consumption expenditures and two variables related to IP (IP:Fuel and IP:final products and nonindustrial supplies).

\begin{table}[H]
	\centering
	\caption{Ratio between MSPE of different approaches and MSPE of the benchmark model (\ref{eq:ar}). For each horizon $h$, the ratios that not exceeds 105\% of the smallest one are labeled as the best and shown in bold.}
	\begin{tabular}{ccccccc}
		\hline
		& FA & FS & SMC & adaLASSO & IHT & HTP \\ 
		\hline
		\multicolumn{7}{l}{{\bf EMP}} \\
		$h$=1 & 0.95 & {\bf 0.90} & {\bf 0.94} & {\bf 0.94} & 0.97 & 0.97 \\ 
		$h$=3 & {\bf 0.68} & 0.74 & {\bf 0.66} & 0.79 & 0.72 & 0.70 \\ 
		$h$=6 & 0.88 & {\bf 0.68} & {\bf 0.68} & 0.76 & 0.71 & {\bf 0.67} \\ 
		$h$=12 & 0.62 & 0.69 & {\bf 0.30} & {\bf 0.29} & 0.85 & 0.78 \\ 
		\hline
		\multicolumn{7}{l}{{\bf IP}} \\
		$h$=1 & 0.99 & {\bf 0.81} & {\bf 0.78} & 0.91 & 0.83 & 0.83 \\ 
		$h$=3 & 0.67 & {\bf 0.63} & 0.85 & 0.84 & 0.83 & 0.83 \\ 
		$h$=6 & 0.93 & 0.87 & 0.89 & 1.24 & {\bf 0.81} & 0.90 \\ 
		$h$=12 & {\bf 1.00} & 1.27 & 1.72 & 1.13 & 1.58 & 1.56 \\ 
		\hline
		\multicolumn{7}{l}{{\bf CPI}} \\
		$h$=1 & {\bf 0.95} & 1.13 & 1.10 & 1.04 & {\bf 0.96} & {\bf 0.98} \\ 
		$h$=3 & 0.98 & 0.88 & 1.09 & {\bf 0.78} & {\bf 0.75} & {\bf 0.76} \\ 
		$h$=6 & 1.05 & 0.82 & 0.91 & 0.84 & {\bf 0.77} & 1.00 \\ 
		$h$=12 & 1.26 & 1.12 & {\bf 0.81} & 1.18 & {\bf 0.81} & 1.14 \\ 
		\hline
	\end{tabular}
	\label{table:real}
\end{table}

In terms of IP, neither FA nor VS approaches improve one-year-ahead forecasting ($h=12$) over AR, which implies other economic variables do not help long-term forecast of IP. While for short-term forecasts ($h=1$, 3 and 6), both FA and VS methods are better than AR, with VS method--FS--performs either the best or close to the best among all the methods. In addition, FS surpass FA by reducing MSPE by 18\%, 6\% and 6\% for 1, 3 and 6-months ahead forecast respectively. Among the predictors selected by FS, IP:durable subcategory, S\&P’s composite common stock:dividend yield and  3-month treasury C minus FEDFUNDS occurs the most often. Some other subcategories of IP are also frequently selected (more than 24 times). This leads us to conclude that, besides its own subcategories, the stock market and interest rate can forecast the movement of IP within six months.

The results of CPI show that, across all horizons, FA does not have significant advantage over AR, and even worse than it for one-year ahead forecast. This finding is consistent with the results in \cite{FREDMD}. However, some of the VS methods do achieve substantial improvement over AR, especially the IHT algorithm. IHT performs the best among all the methods across all horizons, which reduces the MSPE of AR by 25\%, 23\% and 19\% for 3, 6 and 12-months ahead forecasts respectively.  Comparing FS with FA, the improvement is even more apparent: FS reduce the MSPE of FA by 23\%, 27\% and 36\% for 3, 6, and 12-months ahead forecasts respectively. 

The predictors selected by IHT are listed in Table \ref{table:sltvar} in Appendix \ref{sec:appA}. For short-term forecast ($h=1$, 3 and 6), only a handful of predictors are included in the model, and the predictors occurred most are real M2 money stock, $t$-dated CPI and its transportation subcategory, and oil price. This finding is consistent with economic theory and the composition of CPI. First, the money supply M2 has prediction power for CPI. This argument is supported by the classical quantity theory of money \citep{mill1965principles} and evidenced by some empirical studies \citep{Bachmeier2005, BaiNg2008}. Second, since CPI series has temporal dependence, it is natural to use current value ($t$-dated CPI) to forecast its future value. Third, transportation is the second largest category in CPI and is very sensitive to the oil price. Oil price also has direct impact on prices of many industrial materials which are the upstream prices of consumer price. In addition, oil price has great influence on other aspects of US economy such as stock market and investment. All these points make oil price an important index for CPI forecasting.

To summarize, some VS methods can provide better forecasts than FA approach, especially in the forecasting of EMP and CPI where SMC and IHT improve upon FA to a large extent. Moreover, for each target variable, the predictors selected by the best VS method are consistent with the underlying economic theory, which highlights the good interpretability of VS approach.


\section{Conclusion and Discussion}
\label{chap4:conclusion}
FA and VS approach indicate two different directions in economic forecasting. FA approach implies the target variable has many relevant predictors which can be explained by a few latent factors, while VS approach assumes only a handful of predictors have prediction powers on the target variable. Which approach is the best depends on the true data structure and the target variable to be forecasted. This paper aims to draw readers' attention on VS approach, which is less emphasized in the economic literature. In this paper, we introduce several cutting-edge VS algorithms to economic forecasting and compare to FA approach. It turns out for some target variables, VS approach is superior than FA approach. In particular, SMC significantly outperforms FA in forecasting of EMP, FS is superior than FA and AR for short-term forecasting of IP, and IHT is the best when predicting CPI. These methods also provide interpretable models which well explain the relationship between the target variable and the selected predictors. The second contribution of this paper lies on the overview and comparison among four different groups of VS methods. The last two groups, GDS-type algorithms and meta-heuristic algorithms, are popular in computer science but have not been widely used in economic forecasting. Several simulation studies are conducted to compare their prediction performance and variable selection accuracy. The most interesting finding is that, in all the simulation studies, the classical procedure FS works pretty well and sometimes even better than some advanced algorithms. Among all the reviewed VS methods, only the very time-consuming algorithm, SMC, is slightly but not significantly better than FS. 

In economic forecasting, it is often the case that the underlying data structure is very complex. The relationships among different economic variables may vary in different time period. Therefore, it is unrealistic to expect any approach to be uniformly better than others. In the past two decades, several works have been developed to combine the ideas of FA and VS into one path, called supervised factor models (SFM), which includes targeted predictor approach \citep{BaiNg2008}, supervised principle component analysis \citep{bair2006supervisedPCA} and combining forecasts using principal components \citep{tu2019forecasting}. These methods take into consideration the target variable when estimating the latent factors. For example, \cite{BaiNg2008} and \cite{tu2019forecasting} first apply least angle regression to select relevant predictors and then construct factors only within the selected predictors. Most of existing methods only consider the commonly used $l_1$ regularization method, and little attention has been paid on $l_0$ constraint optimization algorithms such as the IHT and SMC algorithms. SFM under $l_0$ constraint optimization is an approach worthy for investigation and will be covered in a future paper.

\section*{Conflict of Interest Statement}
No potential conflict of interest is declared by the authors.

\bibliography{L0}
\newpage
\appendix
\renewcommand\thetable{\thesection.\arabic{table}}
\renewcommand\thefigure{\thesection.\arabic{figure}}
\renewcommand\theequation{\thesection.\arabic{equation}}

\begin{center}
   {\LARGE {\bf Supporting Information}} 
\end{center}

\section{Appendix: Tables and Figures}
\label{sec:appA}
\setcounter{table}{0}
\setcounter{figure}{0}

\begin{table}[H]
	\centering
	\caption{In-sample precision and recall in simulation setting 2. The mean of 100 repetitions is presented, followed by the standard error in the parenthesis.}
	\begin{tabular}{cccccccccc}
		\hline
		\multicolumn{5}{c}{{\bf precision}} & \multicolumn{5}{c}{{\bf recall}} \\\cmidrule(r){1-5}\cmidrule(r){6-10}
		FS &  SMC & adaLASSO & IHT & HTP & FS & SMC & adaLASSO & IHT & HTP \\  
		\multicolumn{10}{l}{$\boldsymbol{T=50}$} \\
		0.72 & 0.70 & 0.11 & 0.61 & 0.59 & 0.22 & 0.31 & 0.74 & 0.21 & 0.20 \\ 
		(0.04) & (0.04) & (0.00) & (0.04) & (0.04) & (0.02) & (0.03) & (0.02) & (0.02) & (0.02) \\
		\multicolumn{10}{l}{$\boldsymbol{T=100}$} \\
		0.93 & 0.97 & 0.11 & 0.67 & 0.73 & 1.00 & 1.00 & 1.00 & 0.75 & 0.80 \\ 
		(0.01) & (0.01) & (0.01) & (0.02) & (0.02) & (0.00) & (0.00) & (0.00) & (0.02) & (0.02) \\
		\multicolumn{10}{l}{$\boldsymbol{T=200}$} \\
		0.99 & 0.98 & 0.14 & 0.98 & 0.99 & 1.00 & 1.00 & 1.00 & 0.99 & 1.00 \\ 
		(0.01) & (0.01) & (0.01) & (0.01) & (0.00) & (0.00) & (0.00) & (0.00) & (0.00) & (0.00) \\ 
		\hline
	\end{tabular}
	\label{table:set2precisionrecall}
\end{table}

\begin{table}[H]
	\centering
	\caption{In-sample precision and recall in simulation setting 3. The mean of 100 repetitions is presented, followed by the standard error in the parenthesis.}
	\begin{tabular}{cccccccccc}
		\hline
		\multicolumn{5}{c}{{\bf precision}} & \multicolumn{5}{c}{{\bf recall}} \\\cmidrule(r){1-5}\cmidrule(r){6-10}
		FS &  SMC & adaLASSO & IHT & HTP & FS & SMC & adaLASSO & IHT & HTP \\  
		\multicolumn{10}{l}{$\boldsymbol{T=50}$} \\
		0.11 & 0.13 & 0.08 & 0.07 & 0.06 & 0.04 & 0.06 & 0.34 & 0.03 & 0.03 \\ 
		(0.02) & (0.02) & (0.00) & (0.01) & (0.01) & (0.01) & (0.01) & (0.01) & (0.01) & (0.01) \\ 
		\multicolumn{10}{l}{$\boldsymbol{T=100}$} \\
		0.42 & 0.55 & 0.12 & 0.21 & 0.19 & 0.29 & 0.4 & 0.66 & 0.16 & 0.14 \\ 
		(0.02) & (0.02) & (0.00) & (0.02) & (0.02) & (0.02) & (0.02) & (0.01) & (0.01) & (0.02) \\ 
		\multicolumn{10}{l}{$\boldsymbol{T=200}$} \\
		0.67 & 0.72 & 0.14 & 0.29 & 0.47 & 0.72 & 0.78 & 0.85 & 0.29 & 0.45 \\ 
		(0.01) & (0.02) & (0.01) & (0.02) & (0.03) & (0.01) & (0.01) & (0.01) & (0.02) & (0.02) \\ 
		\hline
	\end{tabular}
	\label{table:set3precisionrecall}
\end{table}

\begin{table}[H]
\caption{Id of selected predictors for each target variable. The number in the parenthesis counts how many the predictor is selected in the 48 rolling-window forecasts. For each target variable, we report the predictors from the best VS method. The meaning of each id can be found in \cite{FREDMD}. If an id has a ``\_" and a number $l$ attached at the end, it stands for the $l$-lagged value. For example, ``IPFPNSS\_2" means $\text{IPFPNSS}_{t-2}$.}
\centering
\scalebox{0.8}{
	\begin{tabular}{l>{\raggedright\arraybackslash}p{14.8cm}}
		\hline
		\multicolumn{2}{c}{{\bf EMP: predictors selected by SMC}} \\[0.15cm]
		\vspace{0.15cm}
		h=1 & PAYEMS(47) M2SL\_3(40) IPFUELS\_5(40) DPCERA3M086SBEA(25) IPFPNSS\_2(24)  \\
		\vspace{0.15cm}
		h=3 & PAYEMS(48)  \\
		\vspace{0.15cm}
		h=6 & PAYEMS(47) TB3SMFFM\_5(39)  \\
		\vspace{0.15cm}
		h=12 & PAYEMS(48) TB3SMFFM\_2(35) COMPAPFFx(13)  \\
		\hline
		\multicolumn{2}{c}{{\bf IP: predictors selected by FS}} \\[0.15cm]
		\vspace{0.15cm}
		h=1 & IPDMAT\_1(48) S.P.div.yield\_1(38)  \\
		\vspace{0.15cm}
		h=3 & IPDMAT(48) S.P.div.yield(48) TB3SMFFM\_5(46) S.P.div.yield\_1(30) CUMFNS\_1(28) IPMANSICS\_1(26) T10YFFM(25) IPCONGD(24) COMPAPFFx(24) IPDMAT\_1(23) HOUSTNE(18) M2REAL\_3(15) EXUSUKx\_1(14) EXJPUSx(13) AAAFFM(12) T1YFFM\_3(12) NDMANEMP\_4(12)  \\
		\vspace{0.15cm}
		h=6 & IPDMAT(47) TB3SMFFM\_4(31) S.P.div.yield(24) S.P..indust\_2(18) IPMANSICS\_1(16) CUMFNS\_1(16) USWTRADE(14) HOUSTNE(13) S.P.PE.ratio(13) DMANEMP(12) MANEMP\_1(12)  \\
		\hline
		\multicolumn{2}{c}{{\bf CPI: predictors selected by IHT}} \\[0.15cm]
		\vspace{0.15cm}
		h=1 & M2REAL(48) CPITRNSL\_1(48) OILPRICEx(34) IPBUSEQ\_1(29) IPFUELS\_1(28) M2SL\_3(27) TB3MS(19) S.P.div.yield(16) ANDENOx\_5(14)  \\
		\vspace{0.15cm}
		h=3 & M2REAL(48) CPIAUCSL(34) CUSR0000SA0L5(25) CPITRNSL\_1(25) CUSR0000SA0L2(22) CUSR0000SAC\_1(17)  \\
		\vspace{0.15cm}
		h=6 & M2REAL(48) CPIAUCSL(48) CPITRNSL\_1(48) CUSR0000SA0L2(33) CUSR0000SA0L5(29) T1YFFM\_5(24) M2SL(12) DNDGRG3M086SBEA(12)  \\
		h=12 & ACOGNO(48) M2REAL(48) CPIAUCSL(48) CUSR0000SA0L2(43) CUSR0000SA0L5(43) CPITRNSL\_1(37) DNDGRG3M086SBEA(34) IPNMAT\_1(34) ACOGNO\_1(34) ACOGNO\_2(34) CUSR0000SA0L5\_1(33) CPIULFSL\_1(32) CUSR0000SAC\_1(31) CPIULFSL(29) CPIAUCSL\_1(27) CUSR0000SA0L2\_1(27) ISRATIOx(26) CUSR0000SAC(25) IPNMAT(24) M2REAL\_1(24) DNDGRG3M086SBEA\_1(24) CES0600000007\_5(24) ACOGNO\_3(23) CMRMTSPLx(22) EXUSUKx(22) TB3SMFFM\_5(20) RETAILx(19) S.P..indust(19) CPITRNSL(19) IPFUELS\_1(19) RETAILx\_1(18) DPCERA3M086SBEA(17) PCEPI(17) ISRATIOx\_1(16) CPIAUCSL\_2(16) EXUSUKx\_1(15) UMCSENTx\_1(15) CUSR0000SA0L5\_2(15) BUSINVx(14) CMRMTSPLx\_1(14) TB3SMFFM\_4(14) VXOCLSx\_1(13) ACOGNO\_4(13)  \\ 
		\hline
	\end{tabular}
}
	\label{table:sltvar}
\end{table}

\begin{table}[H]
	\centering
	\caption{Ratio between MSPE of different approaches and MSPE of benchmark AR model. For each horizon $h$, the ratios that not exceeds 105\% of the smallest one are labeled as the best and shown in bold. For the data pre-processing, the outliers are handled using strategy 2.}
	\begin{tabular}{ccccccc}
		\hline
		& FA & FS & SMC & adaLASSO & IHT & HTP \\ 
		\hline
		\multicolumn{7}{l}{{\bf EMP}} \\
		$h$=1 & 1.09 & 0.90 & 0.91 & 0.92 & {\bf 0.74} & 0.94 \\ 
        $h$=3 & 0.96 & 0.71 & {\bf 0.69} & 0.81 & {\bf 0.72} & {\bf 0.70} \\ 
        $h$=6 & 1.00 & {\bf 0.68} & {\bf 0.68} & 0.75 & 0.74 & {\bf 0.67} \\ 
        $h$=12 & 0.77 & 0.64 & 0.31 & {\bf 0.29} & 0.79 & 0.66 \\ 
		\hline
		\multicolumn{7}{l}{{\bf IP}} \\
		$h$=1 & 1.07 & {\bf 0.88} & 0.91 & 0.91 & {\bf 0.86} & {\bf 0.85} \\ 
        $h$=3 & 0.87 & {\bf 0.73} & 0.87 & 0.90 & 0.87 & 0.87 \\ 
        $h$=6 & 0.98 & 0.92 & 0.99 & 1.24 & {\bf 0.81} & 0.86 \\ 
        $h$=12 & {\bf 1.01} & 1.52 & 1.51 & 1.13 & 1.59 & 1.52 \\ 
		\hline
		\multicolumn{7}{l}{{\bf CPI}} \\
		$h$=1 & {\bf 0.87} & 1.20 & 1.04 & 0.98 & {\bf 0.91} & 0.99 \\ 
        $h$=3 & 1.17 & 0.91 & 1.03 & 0.81 & {\bf 0.73} & {\bf 0.74} \\ 
        $h$=6 & 1.09 & {\bf 0.82} & 0.97 & {\bf 0.84} & {\bf 0.83} & 1.03 \\ 
        $h$=12 & 1.17 & 1.20 & {\bf 1.03} & 1.08 & {\bf 1.02} & 1.24 \\ 
		\hline
	\end{tabular}
	\label{table:real2}
\end{table}


\section{Appendix: Algorithms}
\label{sec:appB}
\setcounter{equation}{0}

In this section, we describe the detailed algorithms of the following VS methods: iterative hard thresholding (IHT), hard thresholding pursuit (HTP), forward selection (FS), backward elimination (BE) and sequential Monte Carlo (SMC), including their pseudo codes and choices of algorithm parameters. We adopt the same framework and notations as those in the main context to make the description consistent. 

\subsection{Gradient Descent Algorithms with Sparsification (GDS)}
\label{subsec:appGDS}

Recall the objective function of VS problem:
\begin{equation}
\min_{\bbeta}\;\norm{\by - \beta_0\bone - \bbX\bbeta}^2_2 \quad\text{subject to} \quad \norm{\bbeta}_0\leq k,
\label{eq:vsoptdup}
\end{equation}
where $\by$ is $T$-dimensional vector, $\bbX$ is $T\times p$ matrix and $\bbeta$ is $p$-dimensional vector. The $i$th column of matrix $\bbX$, denoted by $\bx_i$, is the vector of observations of the $i$th predictors across time $t$. Our loss function is the sum of squared errors $\norm{\by - \beta_0\bone - \bbX\bbeta}^2_2$, its gradient at any arbitrary $\bbeta$ is $\nabla f(\bbeta)=2\bX'\bX\bbeta-2\bX'\by$. The following two pseudo codes illustrate the IHT and HTP algorithms respectively. 

\begin{algorithm}[H]
	Input initial estimation $\bbeta^{(0)}=\bzero$; iteration r=0; step size $\eta$.\\
	\While{(${R^2}^{(r)}-{R^2}^{(r-1)}>\epsilon_1$ or $\norm{\bbeta^{(r)}-\bbeta^{(r-1)}}_2^2>\epsilon_2$)}{
		{\bf 1. }$\bbeta^{(r+1)}=P_k\lsk \bbeta^{(r)}-\eta\lmk2\bX'\bX\bbeta^{(r)}-2\bX'\by\rmk \rsk$, where operator $P_k(\bv)$ keeps the largest $k$ elements in magnitude of vector $\bv$ and set the rests to zeros.\\
		{\bf 2. }Calculate ${R^2}^{(r+1)}=1 - \norm{\by-\bX\bbeta^{(r+1)}}_2^2/SST$, $\quad r \leftarrow r+1$.
	}
	Output $\bbeta^{(r)}$.
	\caption{\bf IHT algorithm}
\end{algorithm}
\vskip 0.5cm
\begin{algorithm}[H]
	Input $\bbeta^{(0)}=\bzero$; t=0; $\eta$.\\
	\While{(${R^2}^{(r)}-{R^2}^{(r-1)}>\epsilon_1$ or $\norm{\bbeta^{(r)}-\bbeta^{(r-1)}}_2^2>\epsilon_2$)}{
		{\bf 1.} $\tilde{\bbeta}^{(r+1)}=P_k\lsk \bbeta^{(r)}-\eta\lmk2\bX'\bX\bbeta^{(r)}-2\bX'\by\rmk \rsk$. \\
		{\bf 2.} $\bbeta^{(r+1)} = \argmin_{\supp\lsk\bbeta\rsk\subseteq \supp\lsk\tilde{\bbeta}^{(r+1)}\rsk}\norm{\by - \beta_0\bone - \bbX\bbeta}^2_2$.\\
		{\bf 3. }Calculate ${R^2}^{(r+1)}=1 -\norm{\by-\bX\bbeta^{(r+1)}}_2^2/SST$, $\quad r \leftarrow r+1$.
	}
	Output $\bbeta^{(r)}$.
	\caption{\bf HTP algorithm}
\end{algorithm}
In these two algorithms, we set $\epsilon_1=0.005$ and $\epsilon_2=0.01$. The choice of step size $\eta$ is based on the dataset. Too small step size will limit the searching space and cause the algorithms trapped in a local minimum, while the algorithms will not converge with too wide step size. For different dataset, we chose a large step size which still guarantees convergence.

\subsection{Fast Updating Algorithms for FS and BE}
\label{subsec:appclassical}

The VS objective function (\ref{eq:vsoptdup}) has the following equivalent form:
\begin{equation}
\min_{|U|=k}\norm{\by-\hat{\beta}_0\bone-\bbX_U\hbbeta_U}^2_2, \quad \hbbeta_U=\lsk\bbX'_U\bbX_U \rsk^{-1}\bbX'_U\by.
\label{eq:vsoptdup2}
\end{equation}
Here $U \subseteq \{1,\dots,p\}$ is a subset of all predictor indexes, $\bbX_U:=[\bx_i]_{i\in U}$ is the corresponding sub-matrix of $\bbX$ which consists of the predictors in $U$,  and $\bbeta_U=[\beta_i]_{i\in U}$ is the corresponding vector of coefficients. For FS, BE and SMC, we use this objective function (\ref{eq:vsoptdup2}), since it makes the description of these methods more convenient.
The optimal solution of (\ref{eq:vsoptdup2}) is denoted as $U_k$, and its estimate is denoted as $\hat{U}_k$.

\paragraph{Updating algorithm of FS} FS starts with $U$ as empty set. In each step, FS adds on one predictor that gives the best improvement of SSE. The algorithm stops until we have$k$ predictors in the model, and the corresponding set $U$ is the final estimation of $U_k$. 
When adding a new predictor, instead of running the OLS, we can use the following algorithm to avoid the calculation of inverse of matrix $\bbX'_U\bbX_U$. 

Suppose we already have predictors $\bbX_{i\in U}=[\bx_i]_{i\in U}$ in our regression, we add a new predictor $\bz$ and want to obtain the new SSE. Set $\bbZ=\lmk \bbX_U, \bz  \rmk$ is the corresponding new design matrix, $P_{\bbX_U}=\bbX_U(\bbX_U'\bbX_U)^{-1}\bbX_U'$ and $P_{\bbZ}=\bbZ(\bbZ'\bbZ)^{-1}\bbZ'$ are the projection matrix of $\bbX_U$ and $\bbZ$ respectively. After some algebra calculation we get
\begin{equation}
(\bbZ'\bbZ)^{-1}=\begin{bmatrix}
(\bbX_U'\bbX_U)^{-1}+(\bbX_U'\bbX_U)^{-1}\bbX_U'\bz\bz'\bbX_U(\bbX_U'\bbX_U)^{-1}/e_{\bz|\bbX_U} & -(\bbX_U'\bbX_U)^{-1}\bbX_U'\bz/e_{\bz|\bbX_U} \\
\bz'\bbX_U(\bbX_U'\bbX_U)^{-1}/e_{\bz|\bbX_U} & 1/e_{\bz|\bbX_U} \\
\end{bmatrix},
\label{eq:ivZtZ}
\end{equation}
\begin{equation}
P_{\bbZ}=P_{\bbX_U}+\frac{1}{e_{\bz|\bbX_U}}P_{\bbX_U}\bz\bz'P_{\bbX_U} - \frac{2}{e_{\bz|\bbX_U}}\bz\bz'P_{\bbX_U} + \frac{1}{e_{\bz|\bbX_U}}\bz\bz',
\end{equation}
where $e_{\bz|\bbX_U}=\bz'(\bI-P_{\bbX_U})\bz$ is the SSE of $\bz$ regressed on $\bbX_U$. Thus the decrement in SSE after adding $z$ is:
\begin{equation}
    \by'(P_{\bbZ}-P_{\bbX_U})\by=(\by'\be_{\bz})^2/e_{\bz|\bbX_U},
\end{equation}
where $\be_{\bz}=\bz - P_{\bbX_U}\bz$ is the residuals of $\bz$ regressed on $\bbX_U$. In this way we can calculate the corresponding decrement of SSE ($\by'(P_{\bbZ}-P_{\bbX_U})\by$) of adding $\bz$ to the regression without inverting matrix $\bbZ'\bbZ$. To be noticed, if $\bz$ is almost a linear combination of columns in $\bbX_U$ (collinearity), then $P_{\bbZ} \approx P_{\bbX_U}$ and the change in SSE after adding $\bz$ is 0.

\paragraph{Updating algorithm of BE} BE begins with the full model which contains all the predictors, i.e. $U=\{1,\dots,p\}$. In each step, BE deletes one predictor which gives the minimum SSE increment. Similar to FS, there is an efficient way to calculating the inverse of matrix $\bbX_U'\bbX_U$ after deleting the $j$th column of $\bbX_U$ as well as the corresponding increment of SSE. Let  $\bx_j$ be the $j$th column of $\bbX_U$, $\bbtZ$ be the design matrix after deleting the $j$th column and $P_{\bbtZ}$ be the corresponding projection matrix of $\bbtZ$. Based on the equation in (\ref{eq:ivZtZ}), the following algorithm is straightforward.

First, in matrix $(\bbX_U'\bbX_U)^{-1}$, permute its $j$th row and $j$th column to the last row and column. The matrix after permutation is denoted as $\bbM$. Then partition the matrix $\bbM$ as follows:
\begin{equation}
    \bbM=\begin{bmatrix} \bbM_{11} & \bu \\ \bu' & m \\ \end{bmatrix}.
\end{equation}
The inverse of matrix $\bbtZ'\bbtZ$ can be calculated as $(\bbtZ'\bbtZ)^{-1}=\bbM_{11} - \bu\bu'/m$. Therefore, the increment of the SSE after deleting the $j$th column is:
\begin{equation}
\by'(P_{\bbX_U}-P_{\bbtZ})\by=(\by'\bbtZ\bu+m\by'\bx_j)^2/m.
\end{equation}

\subsection{SMC \citep{Duan2019}}
\label{subsec:appSMC}

Set $\bbU_k=\{\text{all the permutations of $k$ members of }\{1,\cdots,p\}\}$, The VS problem (\ref{eq:vsoptdup2}) is equivalent to:
\begin{equation}
\max_{U\in\bbU_k}\; \exp\lbk -\norm{\by-\bX_U\hbbeta_U}^2_2 \rbk.
\label{eq:SMCopt}
\end{equation}
For a given $k$, \cite{Duan2019} assigned a discrete distribution function $f(U)$ on $\bbU$ that is proportional to the exponential of negative SSE:
\begin{equation}
f(U)\propto \exp\lbk -\norm{\by-\bX_U\hbbeta_U}^2_2 \rbk.
\end{equation}
The peak of $f(U)$ corresponds to the optimal permutation in (\ref{eq:SMCopt}). Finding the maximum of (\ref{eq:SMCopt}) is converted to generating a sample suitably representing this distribution function. To accomplish this, SMC starts with an easy-to-sample distribution $f_0(U)$, moving "smoothly" to our target distribution $f(U)$ through a sequence of intermediate artificial distributions:
\begin{equation}
f_j(U)\propto f^{\gamma_j}(U)f^{1-\gamma_j}_0(U),
\end{equation}
where $0=\gamma_0<\gamma_1<\cdots<\gamma_J=1$. $\gamma_0=0$ corresponds to $f_0$ and $\gamma_J=1$ corresponds to target $f$. The solution of (\ref{eq:SMCopt}) is denoted as $\hat{U}_k$. 

The initial sampler, denoted by $I(U\in \bbU_k)$, is already described in the main context. Its distribution $f_0(U)$ is:
\begin{equation}
f_0(\{i_1,i_2,\cdots,i_k\})=q_{i_1}\frac{q_{i_2}}{1-q_{i_1}}\frac{q_{i_3}}{1-q_{i_1}-q_{i_2}}\cdots\frac{q_{i_k}}{1-q_{i_1}-\cdots-q_{i_{k-1}}},\quad \{i_1,i_2,\cdots,i_k\}\in\bbU_k.
\label{f0}
\end{equation}
The corresponding initial sample is denoted as $\{U^{(0)}_m\}_{m=1}^M$ with sample size $M=1000$. We call each element $U$ in the sample as "particle". The following describes the weighting-resampling-support boosting scheme. By iterating this scheme, we can move the sample from $f_0$ sequentially to finally arrive at a representative sample for the target distribution $f$. Each repetition of the three-step scheme is called a round. Based on the sample from the $j$th round, denoted as $\{U^{(j)}_m\}_{m=1}^M\sim f_j$, the weighting-resampling-support boosting scheme will produce a new sample for $f_{j+1}$.

{\bf 1. Weighting: } To choose $\gamma_{j+1}$, set the importance weights for the current sample $\{U^{(j)}_m\}_{m=1}^M$ to be $w^{(j)}_m :=w_{\gamma_{j+1},\gamma_j}(U^{(j)}_m) \propto\lsk \exp\lbk-\|\by-\bX_{U^{(j)}_m}\hbbeta_{U^{(j)}_m}\|^2_2\rbk\middle/f_0(U^{(j)}_m) \rsk^{\gamma_{j+1}-\gamma_j}$. The effective sample size (ESS) implied by the importance weight is calculated as $ESS_j=\frac{\lsk \sum_{m=1}^Mw^{(j)}_m\rsk^2}{\sum_{m=1}^M(w^{(j)}_m)^2}$. Chose $\gamma_{j+1}$ to guarantee $ESS_j\geq \eta M$ where $\eta=1/2$ is pre-specified. 

{\bf 2. Resampling: } Resample $M$ particles from $\{U^{(j)}_m\}_{m=1}^M$ based on the importance weights $\{w^{(j)}_m\}_{m=1}^M$, denoted this new sample by $\{U^{(j+1)}_m\}_{m=1}^M$. Then this new sample has equal weight $1/M$ and follows $f_{j+1}$. 

{\bf 3. Support Boosting: } After resampling, the empirical support (number of distinct members) of $\{U^{(j+1)}_m\}_{m=1}^M$ has shrunk. Boosting the empirical support can be accomplished by several Metropolis-Hastings (MH) moves until the cumulative acceptance rate has reached a target level such as 500\%, or maximum MH moves such as 10 moves.\\
{\bf (1)} For each $U\in\{U^{(j+1)}_m\}_{m=1}^M$, move it to $U^*$ by only replacing a random subset $A\subseteq U$. The current SMC sample $U\in\{U^{(j+1)}_m\}_{m=1}^M$ provide a good basis of proposal. Set $c^{(j+1)}_i$ to be the total count of $x_i$ appearing in $\{U^{(j+1)}_m\}_{m=1}^M$, and then define probability $q^{(j+1)}_i=c^{(j+1)}_i/\sum_{i=1}^pc^{(j+1)}_i$, which reflects relative importance of $x_i$ in current sample. \\
{\bf (2)} Fristly random select $A$ out of $U$. One way for replacing $A$ is sampling $|A|$ predictors from  $\{1,\cdots,p\}\backslash A$ based on $\{q^{(j+1)}_i\}$, another way is based on $\{q_i\}$ defined in the Initialization. This is a MH move from $U$ to $U^*$, the resulted new particle is denoted by $U^*$. Duan used a mixed sampling of these two ways as proposal distributions ($\omega=0.5$): 
\begin{equation}
h^{(\omega)}(U^*|\;U^*\backslash A=U\backslash A)=\omega h(U^*|\;U^*\backslash A=U\backslash A) + (1-\omega)f_0(U^*|\;U^*\backslash A=U\backslash A), 
\end{equation}
\begin{align}
	a &= \min\lsk1, \; \frac{f_{j+1}(U^*)h^{(\omega)}(U|\;U^*\backslash A=U\backslash A)}{f_{j+1}(U)h^{(\omega)}(U^*|\;U^*\backslash A=U\backslash A)} \rsk \nonumber\\
	&= \min\lsk1, \; \frac{\exp\lbk-\gamma_{j+1}\|\by-\bX_{U^*}\hbbeta_{U^*}\| \rbk}{\exp\lbk-\gamma_{j+1}\|\by-\bX_{U}\hbbeta_{U}\| \rbk}  \times   \lbk \frac{f_0(U^*)}{f_0(U)} \rbk^{1-\gamma_{j+1}}  \times  \frac{h^{(\omega)}(U|\;U^*\backslash A=U\backslash A)}{h^{(\omega)}(U^*|\;U^*\backslash A=U\backslash A)} \rsk
\end{align}
For each $U\in\{U^{(j+1)}_m\}_{m=1}^M$, do MH moves until the cumulative acceptance rate reaches a target level, say 500\%. $\{U^{(j+1)}_m\}_{m=1}^M\leftarrow\{U^{*(j+1)}_m\}_{m=1}^M$.

After round $J$, we have obtained the sample for $f$. At last,  we duplicate the sample $\{U^{(J)}_m\}_{m=1}^M$, then do support boosting to reduce the duplicates. The final sample of size $2M$.  $\hat{U}_k$ is the particle in the $J$ rounds which has the smallest SSE.

\paragraph{Adding or Trimming Predictors}
The final SMC solution with $k$ predictors, denoted by $\hat{U}_k$, is a good basis for a new target number $k'$. If $k'<k$, SMC randomly removes $k-k'$ predictors with equal probability. If $k'>k$, we can add $k'-k$ predictors into $\hat{U}_k$ based on the probabilities $\{q_i\}$ in the initialization. Denote this sampler of adding/trimming predictors as $T(U\in\bbU_{k'}; \hat{U}_k)$. The initial sampler of finding optimal $k'$ predictors is:
\begin{equation}
I(U\in \bbU_{k'})=\omega'T(U\in\bbU_{k'}; \hat{U}_k) + (1-\omega')I_0(U\in \bbU_k), \quad \omega'=0.1.
\end{equation}
To be noticed, there are two differences from original weighting-resampling-support boosting scheme due to the change in the initial sampler:\\
{\bf Weighting:} Weight is changed to: $w^{(j)}_m\propto\lsk \exp\lbk-\|\by-\bX_{U^{(j)}_m}\hbbeta_{U^{(j)}_m}\|^2_2\rbk\middle/I(U\in \bbU_{k'}; \{q_i\}) \rsk^{\gamma_{j+1}-\gamma_j}$. \\
{\bf Support Boosting:} In the calculation of acceptance rate, We use $I(U\in \bbU_{k'})$ instead of $I_0(U\in \bbU_k)$ in the proposal $h^{(\omega)}(U^*|\;U^*\backslash A=U\backslash A)$ and calculating acceptance rate.\\

\end{document}